\documentstyle[twoside,epsfig]{article}

\catcode`\@=11
\long\def\@makefntext#1{
\protect\noindent \hbox to 3.2pt {\hskip-.9pt  
$^{{\eightrm\@thefnmark}}$\hfil}#1\hfill}               

\def\thefootnote{\fnsymbol{footnote}}
\def\@makefnmark{\hbox to 0pt{$^{\@thefnmark}$\hss}}    
        
\def\ps@myheadings{\let\@mkboth\@gobbletwo
\def\@oddhead{\hbox{}
\rightmark\hfil\eightrm\thepage}   
\def\@oddfoot{}\def\@evenhead{\eightrm\thepage\hfil
\leftmark\hbox{}}\def\@evenfoot{}
\def\sectionmark##1{}\def\subsectionmark##1{}}
\input epsf

\oddsidemargin=\evensidemargin
\renewcommand{\thefootnote}{\fnsymbol{footnote}}

\newcounter{sectionc}\newcounter{subsectionc}\newcounter{subsubsectionc}
\renewcommand{\section}[1] {\vspace{12pt}\addtocounter{sectionc}{1} 
\setcounter{subsectionc}{0}\setcounter{subsubsectionc}{0}\noindent 
        {\tenbf\thesectionc. #1}\par\vspace{5pt}}
\renewcommand{\subsection}[1] {\vspace{12pt}\addtocounter{subsectionc}{1} 
        \setcounter{subsubsectionc}{0}\noindent 
        {\bf\thesectionc.\thesubsectionc. {\kern1pt \bfit #1}}\par\vspace{5pt}}
\renewcommand{\subsubsection}[1] {\vspace{12pt}\addtocounter{subsubsectionc}{1}
        \noindent{\tenrm\thesectionc.\thesubsectionc.\thesubsubsectionc.
        {\kern1pt \tenit #1}}\par\vspace{5pt}}

\newcounter{appendixc}
\newcounter{subappendixc}[appendixc]
\newcounter{subsubappendixc}[subappendixc]
\renewcommand{\thesubappendixc}{\Alph{appendixc}.\arabic{subappendixc}}
\renewcommand{\thesubsubappendixc}
        {\Alph{appendixc}.\arabic{subappendixc}.\arabic{subsubappendixc}}

\renewcommand{\appendix}[1] {\vspace{12pt}
        \refstepcounter{appendixc}
        \setcounter{figure}{0}
        \setcounter{table}{0}
        \setcounter{lemma}{0}
        \setcounter{theorem}{0}
        \setcounter{corollary}{0}
        \setcounter{definition}{0}
        \setcounter{equation}{0}
        \renewcommand{\thefigure}{\Alph{appendixc}.\arabic{figure}}
        \renewcommand{\thetable}{\Alph{appendixc}.\arabic{table}}
        \renewcommand{\theappendixc}{\Alph{appendixc}}
        \renewcommand{\thelemma}{\Alph{appendixc}.\arabic{lemma}}
        \renewcommand{\thetheorem}{\Alph{appendixc}.\arabic{theorem}}
        \renewcommand{\thedefinition}{\Alph{appendixc}.\arabic{definition}}
        \renewcommand{\thecorollary}{\Alph{appendixc}.\arabic{corollary}}
        \renewcommand{\theequation}{\Alph{appendixc}.\arabic{equation}}
        \noindent{\tenbf Appendix \theappendixc #1}\par\vspace{5pt}}
\newcommand{\subappendix}[1] {\vspace{12pt}
        \refstepcounter{subappendixc}
        \noindent{\bf Appendix \thesubappendixc. {\kern1pt \bfit #1}}
        \par\vspace{5pt}}
\newcommand{\subsubappendix}[1] {\vspace{12pt}
        \refstepcounter{subsubappendixc}
        \noindent{\rm Appendix \thesubsubappendixc. {\kern1pt \tenit #1}}
        \par\vspace{5pt}}

\newcommand{\textlineskip}{\baselineskip=13pt}
\newcommand{\smalllineskip}{\baselineskip=10pt}

\def\eightcirc{
\begin{picture}(0,0)
\put(4.4,1.8){\circle{6.5}}
\end{picture}}
\def\eightcopyright{\eightcirc\kern2.7pt\hbox{\eightrm c}} 

\newcommand{\copyrightheading}[1]
        {\vspace*{-2.5cm}\smalllineskip{\flushleft
        {\footnotesize International Journal of Modern Physics E, #1}\\
        {\footnotesize $\eightcopyright$\, World Scientific Publishing
         Company}\\
         }}

\def\abstracts#1#2#3{{
        \centering{\begin{minipage}{4.5in}\baselineskip=10pt\footnotesize
        \parindent=0pt #1\par 
        \parindent=15pt #2\par
        \parindent=15pt #3
        \end{minipage}}\par}} 

\renewenvironment{thebibliography}[1]
        {\frenchspacing
         \ninerm\baselineskip=11pt
         \begin{list}{\arabic{enumi}.}
        {\usecounter{enumi}\setlength{\parsep}{0pt}     
         \setlength{\leftmargin 12.7pt}{\rightmargin 0pt} 
         \setlength{\itemsep}{0pt} \settowidth
        {\labelwidth}{#1.}\sloppy}}{\end{list}}

\newcounter{itemlistc}
\newcounter{romanlistc}
\newcounter{alphlistc}
\newcounter{arabiclistc}

\newcommand{\fcaption}[1]{
        \refstepcounter{figure}
        \setbox\@tempboxa = \hbox{\footnotesize Fig.~\thefigure. #1}
        \ifdim \wd\@tempboxa > 5in
           {\begin{center}
        \parbox{5in}{\footnotesize\smalllineskip Fig.~\thefigure. #1}
            \end{center}}
        \else
             {\begin{center}
             {\footnotesize Fig.~\thefigure. #1}
              \end{center}}
        \fi}

\newcommand{\tcaption}[1]{
        \refstepcounter{table}
        \setbox\@tempboxa = \hbox{\footnotesize Table~\thetable. #1}
        \ifdim \wd\@tempboxa > 5in
           {\begin{center}
        \parbox{5in}{\footnotesize\smalllineskip Table~\thetable. #1}
            \end{center}}
        \else
             {\begin{center}
             {\footnotesize Table~\thetable. #1}
              \end{center}}
        \fi}

\def\@citex[#1]#2{\if@filesw\immediate\write\@auxout
        {\string\citation{#2}}\fi
\def\@citea{}\@cite{\@for\@citeb:=#2\do
        {\@citea\def\@citea{,}\@ifundefined
        {b@\@citeb}{{\bf ?}\@warning
        {Citation `\@citeb' on page \thepage \space undefined}}
        {\csname b@\@citeb\endcsname}}}{#1}}

\newif\if@cghi
\def\cite{\@cghitrue\@ifnextchar [{\@tempswatrue
        \@citex}{\@tempswafalse\@citex[]}}
\def\citelow{\@cghifalse\@ifnextchar [{\@tempswatrue
        \@citex}{\@tempswafalse\@citex[]}}
\def\@cite#1#2{{$\null^{#1}$\if@tempswa\typeout
        {IJCGA warning: optional citation argument 
        ignored: `#2'} \fi}}

\def\pmb#1{\setbox0=\hbox{#1}
        \kern-.025em\copy0\kern-\wd0
        \kern.05em\copy0\kern-\wd0
        \kern-.025em\raise.0433em\box0}

\def\fnt#1#2{\footnotetext{\kern-.3em
        {$^{\mbox{\scriptsize #1}}$}{#2}}}

\def\fpage#1{\begingroup
\voffset=.3in
\thispagestyle{empty}\begin{table}[b]\centerline{\footnotesize #1}
        \end{table}\endgroup}

\def\runninghead#1#2{\pagestyle{myheadings}
\markboth{{\protect\footnotesize\it{\quad #1}}\hfill}
{\hfill{\protect\footnotesize\it{#2\quad}}}}
\font\tenrm=cmr10
\font\tenit=cmti10 
\font\tenbf=cmbx10
\font\bfit=cmbxti10 at 10pt
\font\ninerm=cmr9

\font\eightrm=cmr8

\def\qed{\hbox{${\vcenter{\vbox{                        
   \hrule height 0.4pt\hbox{\vrule width 0.4pt height 6pt
   \kern5pt\vrule width 0.4pt}\hrule height 0.4pt}}}$}}

\renewcommand{\thefootnote}{\fnsymbol{footnote}}        

\def\bsc{{\sc a\kern-6.4pt\sc a\kern-6.4pt\sc a}}       
\def\bflatex{\bf L\kern-.30em\raise.3ex\hbox{\bsc}\kern-.14em 
T\kern-.1667em\lower.7ex\hbox{E}\kern-.125em X} 

\begin{document}

\runninghead{Nucleon Electromagnetic Form Factors} 
{Nucleon Electromagnetic Form Factors}

\normalsize\textlineskip
\thispagestyle{empty}
\setcounter{page}{1}

\copyrightheading{}                     

\vspace*{0.88truein}

\fpage{1}
\centerline{\bf Nucleon Electromagnetic Form Factors}
\vspace*{0.37truein}
\centerline{\bf Haiyan Gao}
\vspace*{0.37truein}
\centerline{\footnotesize\it  Triangle University Nuclear Laboratory and 
Department of Physics, Duke University}
\baselineskip=10pt
\centerline{\footnotesize\it Durham, North Carolina 27708, U.S.A.}
\centerline{\footnotesize\it Laboratory for Nuclear Science and 
Department of Physics, Massachusetts Institute of Technology}
\baselineskip=10pt
\centerline{\footnotesize\it Cambridge, Massachusetts 02139, U.S.A.}
\baselineskip=10pt
\centerline{\footnotesize\it gao@tunl.duke.edu}

\vspace*{10pt}
\vspace*{0.225truein}

\vspace*{0.21truein}
\abstracts{Nucleon electromagnetic form factors are fundamental 
quantities related to
the charge and magnetization distributions inside the nucleon. 
Understanding the nucleon electromagnetic structure in terms of the underlying 
quark and gluon degrees of freedom of 
quantum chromodynamics is a challenging and urgent task. 
The nucleon electromagnetic form factors have been 
studied in the past extensively from unpolarized electron scattering 
experiments. With the development in polarized beam, recoil polarimetry, 
and polarized target 
technologies, polarization experiments have provided more precise 
data on these quantities.
At the same time, significant theoretical progress in areas 
ranging from effective field theories to lattice QCD 
calculations, has been made in describing these data.
In this article, I review recent experimental and theoretical 
progress on this subject. I will also provide future 
outlook on this topic.}{}{}


\vspace*{1pt}\textlineskip      

\vspace*{-0.5pt}
\noindent

\section{Introduction}

Nucleons (protons and neutrons) are fundamental building blocks of
matter. Nevertheless, they are not fundamental particles.
Otto Stern received the Nobel Prize in Physics in 1943,
for his experimental methods of studying the magnetic properties of
nuclei, in particular for measuring the magnetic moment of the proton
itself. The anomalous magnetic moments of the proton and neutron 
reveal that nucleons are not point-like Dirac particles, 
but particles with underlying structure. 

The electromagnetic form factors of the nucleon have been a
longstanding subject of interest in nuclear and particle physics.
They are fundamental quantities describing
the distribution of charge and magnetization within nucleons.
Probing the nucleon electromagnetic structure has been an ongoing
experimental endeavor since the discovery of the
anomalous magnetic moment of the proton.
 
Electron scattering has been proven to be a very useful tool in probing 
the structure of nucleon and nuclei. The leptonic part of the vertex 
is well understood from Quantum Electrodynamics (QED), thus it is a clean 
probe of hadronic structure. Furthermore, the electromagnetic 
coupling constant is relatively weak, so higher order diagrams are suppressed 
compared to the lowest order one-photon-exchange diagram.
The study of the electromagnetic structure of the proton from
electron-proton elastic scattering was
pioneered by Hofstadter and colleagues at the Stanford Linear
Accelerator Center (SLAC) in the 1950s, for which Hoftstadter received
the Nobel Prize in Physics in 1961. 
The deep-inelastic scattering experiments of electrons from
protons carried out by Friedman, Kendall and Taylor which established
the underlying quark structure of the proton at SLAC, led to the Nobel
Prize in Physics in 1990.

Quantum Chromodynamics (QCD) is the theory of strong interaction in terms
of quark and gluon degrees of freedom. While QCD has been extremely well
tested in the high energy regime, where perturbative QCD is applicable,
understanding confinement and hadron structure in the non-perturbative
region of QCD remains challenging.
Knowledge of the internal structure of protons and neutrons in
terms of quark and gluon degrees of freedom is not only essential for testing 
QCD in the confinement regime, but it also provides a basis for 
understanding more complex, strongly interacting matter 
at the level of quarks and gluons.

The rest of the paper is organized as follows: 
Section II gives an introduction on the nucleon electromagnetic form factors;
Section III contains a discussion on form factor data from unpolarized 
electron scattering experiments; Section IV presents
recent data on the nucleon electromagnetic form factors from 
double-polarization experiments; 
Section V reviews theoretical progress on the nucleon 
electromagnetic form factors; and Section VI provides
future outlook on this subject.

\section{The Nucleon Electromagnetic Form Factors}


From QED, the lowest-order amplitude (Fig.~1) for 
electron-nucleon elastic scattering is given by

\begin{equation}
T_{fi} = - i \int{j_{\mu} ({\frac{-1}{q^2}})J^{\mu}d^{4}x},
\end{equation}

where $q =p'-p$ and the electron transition current is

\begin{equation}
j^{\mu} = -e \bar{u}(k')\gamma^{\mu}u(k)e^{i(k'-k)\cdot x}
\end{equation} 

The nucleon is an extended spin-${\frac{1}{2}}$ object, thus the nucleon 
transition current is more complicated than that of the electron. 
Based on the requirements of covariance under the improper Lorentz group,
current conservation and parity conservation, the nucleon transition current
is written as
\begin{equation}
J^{\mu} = e\bar{u}(p')\left[F_{1}(q^2)\gamma^{\mu} + {\frac{\kappa}{2M}}F_{2}(q^2)i
\sigma^{\mu\nu}q_{\nu}\right]u(p)e^{i(p'-p)\cdot x},
\end{equation}

where $F_{1}$ and $F_{2}$ are two independent form factors, also called 
the Dirac and Pauli form factors, respectively, $\kappa$ is the 
anomalous magnetic moment, and $M$ is the nucleon mass.

Ernst, Sachs, and Wali~\cite{esw} defined the following, the so-called Sachs'
form factors: $G_E(q^2)$ (electric form factor) and $G_M(q^2)$ 
(magnetic form factor), which are written as linear combinations 
of $F_{1}$ and $F_{2}$:
\begin{equation}
G_E = F_1 + {\frac{\kappa q^2}{4M^2}}F_2,
\end{equation}
\begin{equation}
G_M= F_1 + \kappa F_2.
\end{equation}

\begin{figure}[htbp]
\centerline{\epsfig{file=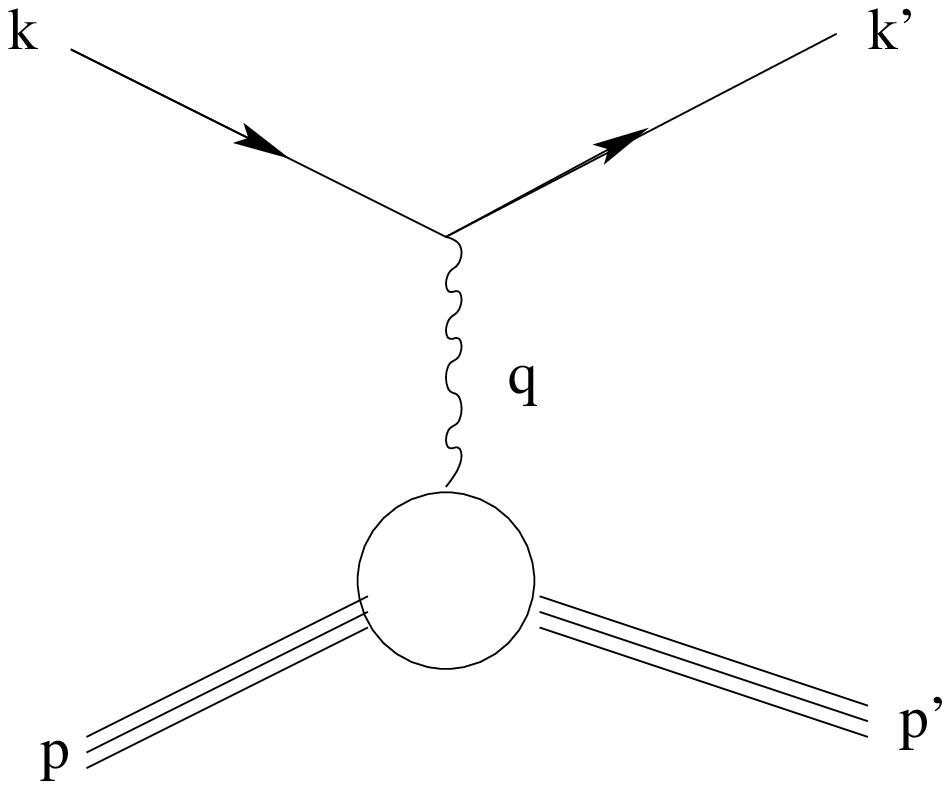,height=3.0in,width=3.0in}}
\fcaption{{\footnotesize\it The one-photon-exchange diagram for 
electron-nucleon elastic scattering.}}
\label{fig:born}
\end{figure}

The Sachs' form factors, $G_E$ and $G_M$, have more intuitive physical 
interpretations than $F_1$ and $F_2$.
In electron scattering, the inverse of the four-momentum 
transfer squared, $Q^2$
($Q^2=-q^2 \ge 0$) is commonly used.
In the limit of $Q^2=0$, 
$$ G^{p}_E(Q^2=0)= 1, G^{n}_E(Q^2=0)= 0$$ and  
$$G^{p}_M(Q^2=0)=\mu_{p}, G^{n}_M(Q^2=0)=\mu_{n},$$
corresponding to the charge and the magnetic moment of the nucleon, 
respectively.
Sachs~\cite{sach} demonstrated that $G_E$ and $G_M$ can be interpreted 
as Fourier transforms of spatial distributions of charge and 
magnetization of the nucleon in the so-called 
Breit frame. For elastic electron-nucleon scattering, 
the Breit frame coincides with 
the center-of-mass frame of the electron-nucleon system. In this reference 
frame, the incoming electron has a momentum of ${\frac{\vec{q}}{2}}$ and
the nucleon initial momentum is $-{\frac{\vec{q}}{2}}$; the scattered electron
has a momentum of $-{\frac{\vec{q}}{2}}$ and the recoil proton has a momentum
of ${\frac{\vec{q}}{2}}$.
Thus, the Breit frame is a special Lorentz frame 
in which $q^2 = - {\vec{q}}^2$, i.e., no energy transfer is involved in this 
particular reference frame. Thus, for each $q^2$ value, there is a Breit frame 
in which the form factors are represented as $G_{E,M}({\vec{q}}^2) =
G_{E,M}(q^2)$, where $G_{E,M}(q^2)$ is determined in the lab frame.
One can therefore perform a three-dimensional Fourier transformation 
once the form factor information is available,
\begin{equation}
\rho(\vec{r}) = \int {\frac{d^{3}q}{{2\pi}^3}} e^{-i \vec{q}\cdot \vec{r}}
{\frac{M}{E(\vec{q})}} G_{E}({\vec{q}}^{2}).
\end{equation}
Although relativity obscures the physical interpretation of such a quantity,
it is analogous to the ``classical'' charge density 
distribution.

In the low $q^2$ region, below the two-pion production threshold
(i.e., $q^2 < t_{0}$, where $t_{0} = (2 m_{\pi})^2 < 2 $ fm$^{-2}$) the energy
transfer in the scattering process is negligible and nucleon electric 
and magnetic form factors
can be taken as the Fourier transforms of the charge and magnetization
radial distributions $\rho_{\rm chg}(r)$ and $\rho_{\rm mag}(r)$ in 
the rest frame of the nucleon.

In the non-relativistic limit, the root-mean-square (rms) charge radius 
of the nucleon is related to the nucleon electric form factor as: 
\begin{equation}
{\frac{<r^2>}{6}} = - {\frac{d{G_{E}(Q^2)}}
{d{Q^2}}}|_{Q^2=0} \>,
\end{equation} 
with the boundary condition $G_{E} (Q^{2} = 0 ) = 1$ for the proton 
and 0 for the neutron.
The corresponding rms magnetic radius is
\begin{equation}
{\frac{<r^2>}{6}} = - {\frac{1}{\mu_N}}{\frac{d{G_{M}(Q^2)}}
{d{Q^2}}}|_{Q^2=0} \>,
\end{equation} 
where $\mu_N$ is the nucleon magnetic moment, $\mu_p$ =2.79 and 
$\mu_n$ =-1.91, in the units of the nucleon magneton.

Recent results from lattice QCD calculations~\cite{dong,thomas} suggest that the 
nucleon rms charge radius can be calculated from first
principles with an uncertainty of only a few percent, and this field
is rapidly evolving due to both improvements in computer architecture
and new algorithms. 
Thus, precise information on this fundamental quantity is essential in terms
of testing the QCD prediction from the lattice.

Accurate information about the proton charge radius is also essential
in conducting high-precision tests of QED
from hydrogen Lamb shift measurements. The standard Lamb shift
measurement probes the 1057 MHz fine structure transition between the
$2S_{1/2}$ and $2P_{1/2}$ states in hydrogen. The hydrogen Lamb shift
can be calculated to high precision from QED using higher order
corrections. The proton rms charge radius is an important input in calculating
the hadronic contribution to the hydrogen Lamb shift. 
 
The two most precise and widely cited determinations of 
the proton charge radius from electron scattering experiments
in the literature give $r_{p} = 0.805(11)$ fm~\cite{hand} and $ r_{p} =
0.862(12)$ fm~\cite{simon}, respectively, 
differing from each other by more than $7\%$. 
While the recent precision hydrogen Lamb shift 
measurements~\cite{weitz,hagley,berkeland,98,can} are in better agreement
with the QED predictions using the smaller value of the proton charge radius
without the two-loop binding effects, they are consistent with the 
larger value of the proton charge radius when two-loop binding effects 
are included in the QED calculations. 
Mostly recently, a QED calculation with
three-loop binding effects has been carried out~\cite{threeloop}.
Before accurate comparisons between theory and experiment can be made, 
either in QCD or QED, a new, high precision experiment 
on the proton charge radius is urgently needed. 

The mean squared charge radius of the neutron
is $<r^2_n> = -0.113 \pm 0.003 \pm 0.004 $ fm$^2$, obtained most
recently from an experiment carried out at the Oak Ridge 
National Laboratory~\cite{neutronradius} in which the neutron transmission 
through a thorogenic liquid $^{208}$Pb has been measured in the neutron
energy range between 0.1 and 360 eV. 

Among the four nucleon electromagnetic form factors, the neutron 
electric form factor, $G^n_E$ is of particular interest, but it is 
also least known due to the lack of free neutron targets
and the smallness of $G^{n}_E$ in general. Its interpretation within many 
models has been obscured by relativistic effects. Recently, 
Isgur~\cite{isgur} 
demonstrated to leading order in the relativistic expansion of a constituent 
quark model that the naive interpretation of $G^n_E$ as 
arising from the neutron's rest frame charge distribution holds due to
a cancellation of the Foldy term against a contribution to the Dirac
form factor, $F_{1}$. 
Enormous experimental progress has been made 
on the nucleon electromagnetic form factors in the last decade or so. 
In particular, new data with significantly improved precision 
from double-polarization 
experiments are available due to the recent advances in 
polarized target, beam and recoil polarimeter technologies. 
More precise data from the next generation
of double-polarization experiments are 
anticipated in the next several years, and these data will allow more 
stringent
tests of theoretical descriptions of the nucleon electromagnetic structure. 
In the next two sections, the experimental progress on the  
nucleon electromagnetic form factors will be reviewed, in particular
the most recent measurements from double-polarization experiments.

\section{Unpolarized Electron Scattering and Nucleon Electromagnetic 
Form Factors}

\subsection{Elastic Electron-Proton Scattering}

In terms of $G_E$ 
and $G_M$, the differential cross-section for electron-nucleon scattering can
be written in the one-photon-exchange picture as:

\begin{equation}
{\frac{d\sigma}{d\Omega}}_{lab} = {\frac{\alpha^2}{4E^2 \sin^{4}{
{\frac{\theta}{2}}}}}
{\frac{E'}{E}}\left({\frac{G^2_{E} + \tau G^{2}_M}{1+\tau}}
\cos^{2}{\frac{\theta}{2}} 
+ 2 \tau G^{2}_M \sin^{2}{{\frac{\theta}{2}}}\right),
\label{eq:unpol}
\end{equation}
where $E'$ and $\theta$ are the scattered electron 
energy and angle, respectively; $\alpha$ is the fine
structure constant, and $\tau = {\frac{Q^2}{4M^2}}$. 

The proton electric ($G^p_E$) and magnetic ($G^p_M$) form factors 
have been studied extensively in the
past from unpolarized electron-proton ($ep$) elastic scattering using the  
Rosenbluth separation technique~\cite{rosenbluth}. 
Eqn.~\ref{eq:unpol} can be re-written as:

\begin{eqnarray}
{\frac{d\sigma}{d\Omega}} &=& {\frac{\alpha^2E'\cos^2{\frac{\theta}{2}}}
{4E^3\sin^4{\frac{\theta}{2}}}}\left[{G^p_E}^2 + 
{\frac{\tau}{\epsilon}}{G^p_M}^2\right]\left({\frac{1}{1+\tau}}\right) \nonumber\\
&=& \sigma_{M}\left[{G^p_E}^2 + 
{\frac{\tau}{\epsilon}}{G^p_M}^2\right]\left({\frac{1}{1+\tau}}\right),
\end{eqnarray}
where $\epsilon={(1+2(1+\tau)\tan^{2}{\frac{\theta}{2}})}^{-1}$ 
is the virtual photon longitudinal polarization, and
$\sigma_{M}$ is the Mott cross section describing the scattering
from a pointlike target:

\begin{equation}
\sigma_{M} = {\frac{\alpha^2 E' \cos^{2}{\frac{\theta}{2}}}{4 E^{3}
    \sin^{4}{\frac{\theta}{2}}}}.
\end{equation}

\vspace{0.8in}
\begin{figure}[htbp]
\centerline{\epsfig{file=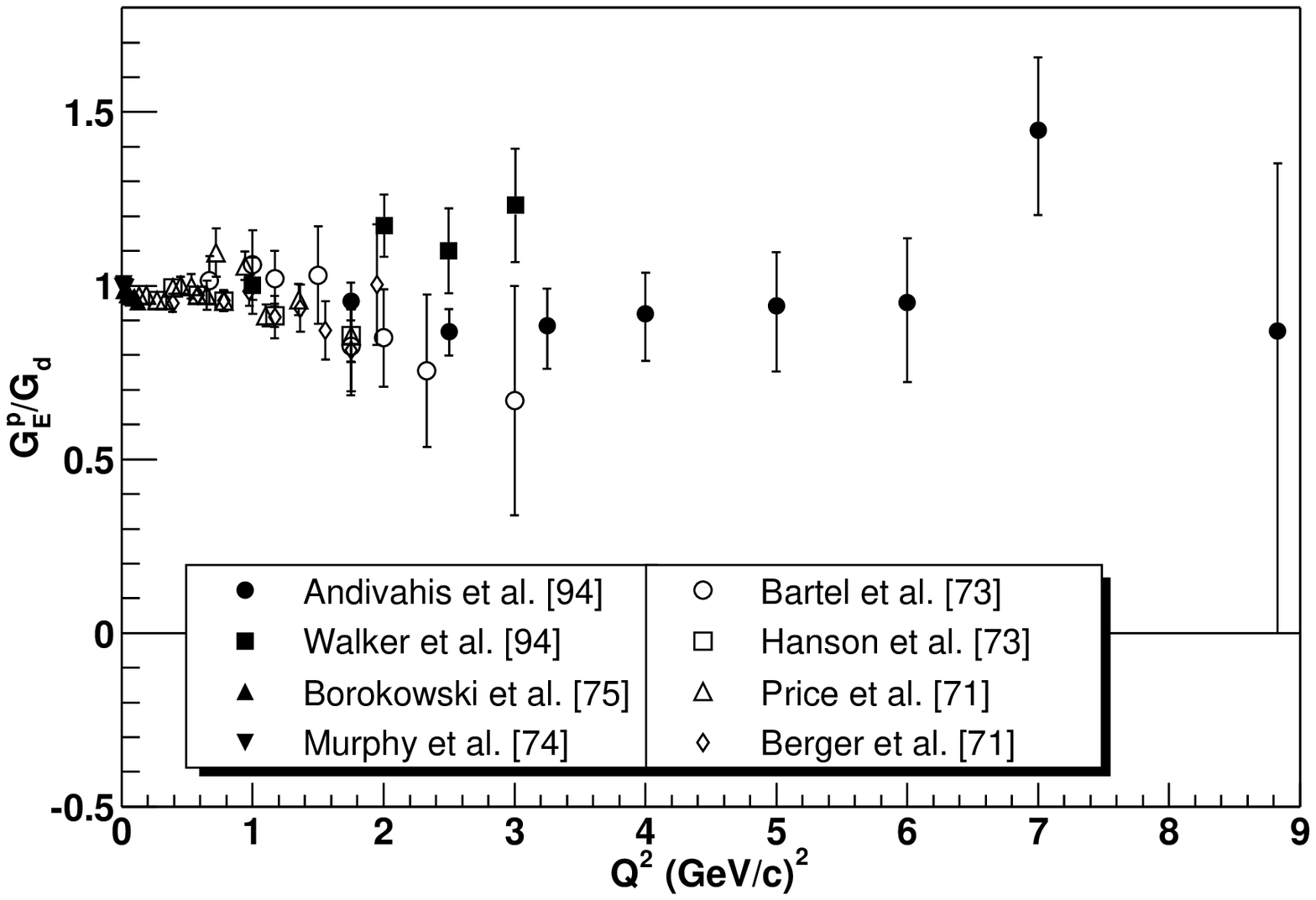,height=6.0cm,width=12.0cm}}
\fcaption{{\footnotesize\it World data published since 1970 on the proton 
electric form factor obtained by the Rosenbluth method from the 
unpolarized cross-section measurements. The data are plotted 
as ${\frac{G^{p}_{E}}{G_D}}$, where $G_D$ is the standard 
dipole parameterization.}}
\label{fig:worlddatagep}
\end{figure}

\begin{figure}[htbp]
\vspace*{13pt}
\centerline{\epsfig{file=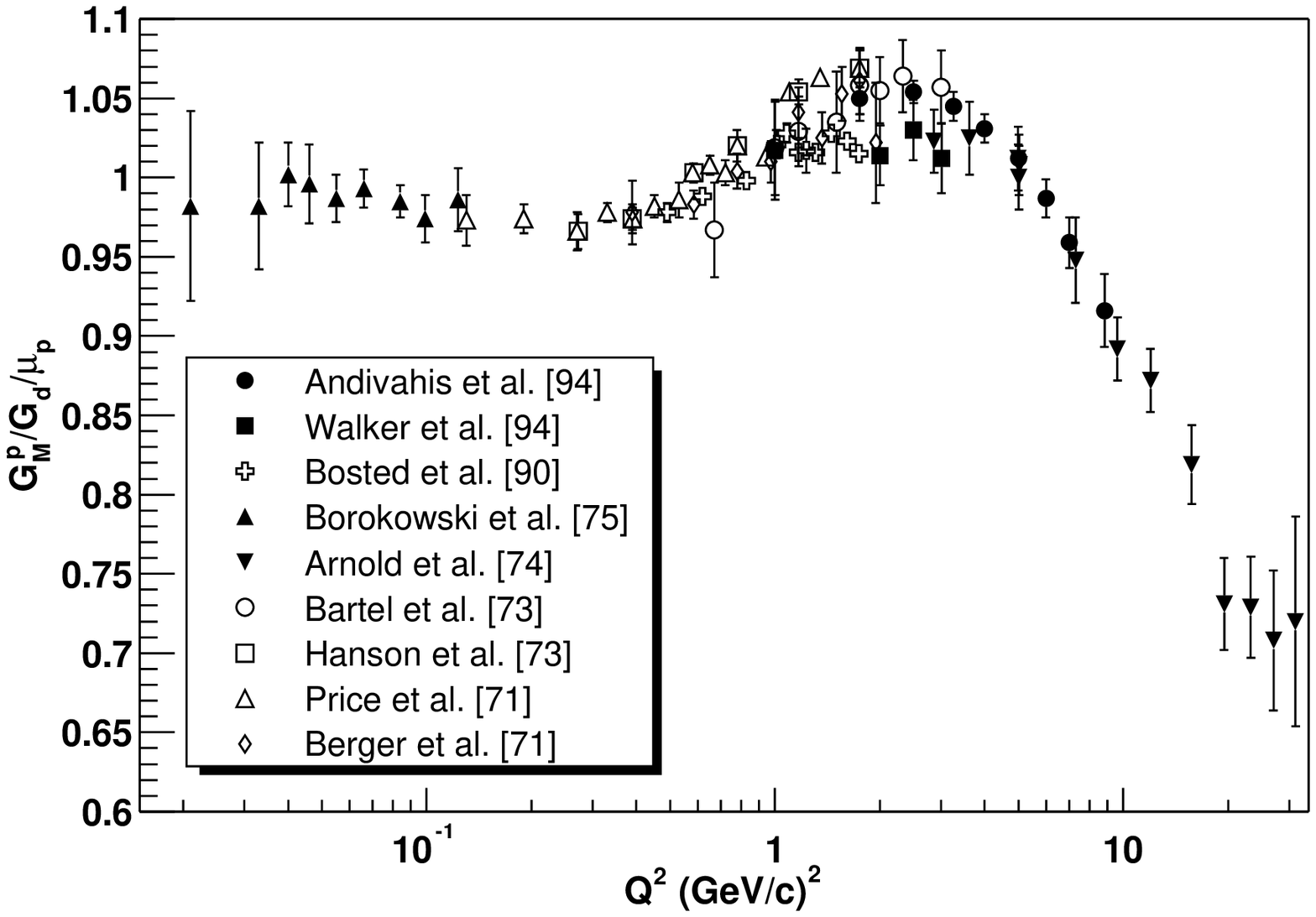,height=6.0cm,width=12.0cm}}
\fcaption{{\footnotesize\it World data published since 1970 on the proton 
magnetic form factor obtained using the Rosenbluth method from unpolarized 
cross-section measurements. The data are plotted 
as ${\frac{G^{p}_{E}}{\mu_{p} G_D}}$, in the unit of the 
standard dipole parameterization.}}
\label{fig:worlddatagmp}
\end{figure}

In the Rosenbluth method, the separation
of ${G^p_E}^2$ and ${G^p_M}^2$ is achieved by measuring the cross section
at a given $Q^2$ value by varying the incident electron beam energy and 
the electron scattering angle. The measured differential cross section is then
plotted as a function of ${\tan^{2}{\frac{\theta}{2}}}$, and one can extract 
information on ${G^p_E}^2$ and ${G^p_M}^2$ from the slope 
and the intercept of the plotted curve.  
While the ${G^p_E}^2$ term dominates the cross section in the 
low $Q^2$ region, the ${G^p_M}^2$ term dominates at large $Q^2$ values. 
Thus, the extraction of $G^p_M$ at low $Q^2$, and $G^p_E$
at large $Q^2$ values becomes difficult using the Rosenbluth technique.
Fig.~\ref{fig:worlddatagep} and Fig.~\ref{fig:worlddatagmp} 
show the world data since 1970~\cite{protondata} on the proton electric 
and magnetic form factors, 
as a function of Q$^2$ using the Rosenbluth separation technique.
The data are shown as
${\frac{G^{p}_{E}}{G_D}}$, and ${\frac{G^{p}_{M}}{\mu_{p} G_D}}$, respectively,
where $\mu_p$ is the proton magnetic moment, and $G_D= (1+Q^2/0.71)^{-2}$ 
is the standard dipole parameterization, and $Q^2$ is in (GeV/c)$^2$.

\subsection{Electron-Deuteron Scattering}

Because of the lack of free neutron targets, little is known about 
the neutron electromagnetic structure. 
The best known quantity is the neutron rms charge radius which is obtained
from thermal neutron scattering experiments.
The neutron electromagnetic form factors 
are known with much less precision than the proton
electric and magnetic form factors. 
They have been deduced in the past from
elastic or quasielastic electron-deuteron scattering. This procedure involves
considerable model dependence. Another complication  
arises from the fact that the net charge of the neutron is zero. As such the 
neutron electric form factor $G^{n}_{E}$ is much smaller 
than its magnetic form factor $G^n_{M}$. Therefore, the magnetic part of the 
contribution dominates the cross section,
which makes it very difficult to extract $G^n_{E}$ from 
unpolarized cross section measurements using deuterium targets.

\vspace{0.7in}

\subsubsection{The Elastic Electron-Deuteron Scattering}

The cross section for the unpolarized elastic electron-deuteron scattering
in the one-photon-exchange approximation is described by the
Rosenbluth formula,

\begin{equation}
{\frac{d\sigma}{d\Omega}}= \sigma_{M}\left[A(Q^2) +
B(Q^2)\tan^{2}{\frac{\theta}{2}}\right].
\end{equation}
The two structure functions: $A(Q^2)$ and $B(Q^2)$
can be separated by the Rosenbluth separation technique discussed previously.
The deuteron is a spin-1 nucleus and the characterization of its charge and 
magnetization distribution requires three form factors: $F_{C}(Q^2)$,
$F_{Q}(Q^2)$, $F_{M}(Q^2)$, the charge monopole, quadrupole and the magnetic 
dipole form factor, respectively.
The structure functions $A(Q^2)$ and $B(Q^2)$ can be expressed in
terms of these three form factors as

\begin{equation}
A(Q^2) = F^{2}_{C}(Q^2) + {\frac{8}{9}}\tau^2 F^{2}_Q(Q^2) + 
{\frac{2}{3}}\tau F^{2}_{M}(Q^2),
\end{equation}

\begin{equation}
B(Q^2) = {\frac{4}{3}}\tau(1+\tau)F^{2}_{M}(Q^2),
\end{equation}
where $\tau={\frac{Q^2}{4M^{2}_{d}}}$.
Thus, it is not possible to separate all three form factors of the
deuteron from the unpolarized elastic electron-deuteron cross section
measurement alone. An additional measurement which involves polarization is
necessary to separate the charge monopole and the charge quadrupole form
factors. Such a polarization experiment
can be either a deuteron tensor polarization measurement 
employing a polarimeter 
by using an unpolarized electron beam and an unpolarized deuteron target 
or an analyzing power measurement by using a
polarized deuteron target and an unpolarized electron beam.

The tensor moment $t_{20}$ is particularly interesting due to its sensitivity
to $F_C$. It is defined as
\begin{equation}
t_{20} = (\frac{1}{\sqrt{2}})<3S_{z}^2 - 2> \\
\vspace{1ex}
=\frac{1}{\sqrt{2}}\left[ \frac{N_{+} + N_{-} -
    2N_{0}}{N_{+}+N_{-}+N_{0}} 
\right], 
\end{equation}
where $N_{+}$, $N_{-}$, and $N_{0}$ are the occupation numbers for the
deuteron magnetic sub-state $M_{s}=1, 0, -1$, respectively.
The deuteron tensor moment $t_{20}$ can be expressed in terms of
the deuteron form factors as:

\begin{equation}
t_{20} = \frac{-1}{\sqrt{2} I_{0}}\left[\frac{8}{3}\tau F_{C} F_{Q}
+ \frac{8}{9}\tau ^2 F_{Q}^2 + \frac{1}{3}\tau 
(1 + 2(1+\tau)\tan^2{\frac{\theta}{2}}) F_{M}^2 \right],
\end{equation}
where $ I_{0} = A(Q^2) + B(Q^2)\tan^2{\frac{\theta}{2}}$.
Therefore, by combining the structure functions $A(Q^2)$, $B(Q^2)$ from the
unpolarized cross section measurement, and the deuteron tensor
moment
$t_{20}$ measurement, one can separate all three deuteron form factors.

\begin{figure}[htbp]
\centerline{\epsfig{file=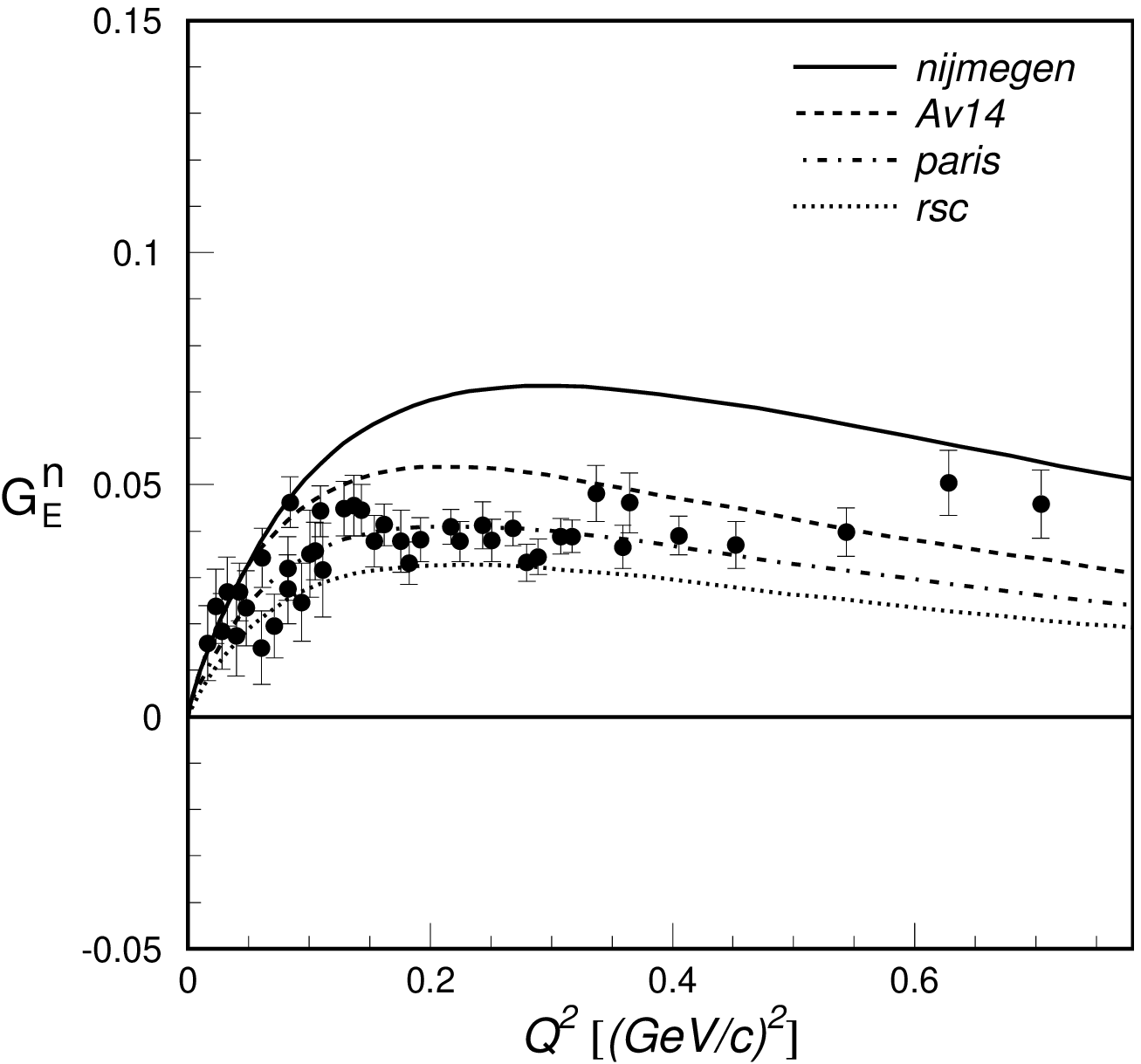,height=5.0cm,width=10.0cm}}
\fcaption{{\footnotesize\it The electric form factor of 
the neutron, $G^n_E$, as a function of four-momentum transfer squared 
from the unpolarized electron-deuteron elastic scattering measurement by 
Platchkov {\it et al.}~\protect\cite{platchkov}. }}
\label{fig:plat}
\end{figure}

The structure function $A(Q^2)$ provides one of the few methods to 
infer the neutron electric form factor, especially in the low $Q^2$ region 
(less than 1.0 (GeV/c)$^2$) where theoretical descriptions of $A(Q^2)$ 
including relativity, meson-exchange currents (MEC), etc. are 
under better control compared to higher $Q^2$ region.
The most systematic information on $G^n_{E}$ at low $Q^{2}$, 
prior to any polarization
experiment, is from the $A(Q^2)$ structure function determined from the
elastic electron-deuteron scattering experiment by
Platchkov {\it et al.}~\cite{platchkov}.
However,  
the extraction procedure is quite complicated. First, 
the subtraction of $F^{2}_{M}(Q^2)$ from $A(Q^2)$ using data on 
$B(Q^2)$ is performed to obtain the corrected $A(Q^2)$ which contains
contributions from $F_{C}(Q^2)$, and $F_{Q}(Q^2)$ only. 
Second, the relativistic and MEC corrections are 
applied to the corrected $A(Q^2)$ to obtain the corresponding $A(Q^2)$
in the impulse picture.
Next, the deuteron structure is removed 
to obtain the nucleon iso-scalar charge
form factor. Finally, the proton electric form factor is subtracted from 
the nucleon iso-scalar charge form factor and $G^{n}_E$ is obtained.
The extracted $G^n_{E}$ values are extremely sensitive to the deuteron 
structure. Fig.~\ref{fig:plat} shows the
$G^n_{E}$ values extracted with the Paris nucleon-nucleon 
potential~\cite{paris} together with a fit of the data (dash-dotted curve). 
Results from fitting the $G^n_{E}$ data extracted with the 
Nijmegen potential~\cite{nijmegen}, 
the Argonne V14 (AV14)~\cite{av14} and the Reid-Soft Core (RSC)~\cite{rsc}
NN potentials are shown as solid, dashed and dotted curves, respectively. 
The large spread represents the uncertainty of $G^n_{E}$ due to the deuteron 
structure, and the absolute scale of 
$G^n_{E}$ contains a systematic uncertainty of about $50\%$ from such an 
extraction.

\vspace{1.7in}

Recently, Schivilla and Sick~\cite{rocco} extracted $G^{n}_E$ from an analysis 
of the deuteron quadrupole form factor $F_{Q}(Q^2)$ data. 
Such an approach is different from the analysis 
employing the structure function $A(Q^2)$ discussed above, 
in which both the charge 
monopole and quadrupole deuteron form factors contribute.
State-of-the-art deuteron calculations based
on a variety of different model interactions and currents show that the 
$F_{Q}$ form 
factor is relatively insensitive to the uncertain two-body operators 
of shorter range because
the long-range one-pion exchange operator dominates 
the two-body contribution to $F_{Q}$.
Also, data on the $F_{Q}$ form factor has improved tremendously with the measurements of the
polarization observable $t_{20}$ in electron-deuteron elastic scattering.
As such, the neutron electric form factor has been extracted 
up to a $Q^2$ value above 40 fm$^{-2}$ (1.55 (GeV/c)$^2$)
from $F_{Q}(Q^2)$ without undue systematic uncertainties from theory. 
These extracted values
of $G^{n}_E$ will be presented later together with $G^{n}_{E}$ 
data from double-polarization experiments.

\subsubsection{Quasielastic Electron-Deuteron Scattering}

Quasielastic electron-deuteron scattering, 
in which the kinematics of the electron scattering
from the nucleon inside the deuteron is selected, 
is the other process involving the deuteron 
which has been used extensively in probing the electromagnetic 
structure of the neutron. It includes both inclusive measurements, 
in which only 
scattered electrons are detected at the quasielastic kinematics, 
and coincidence measurements where both the scattered electron
and the knockout neutron are measured.

The missing mass squared for quasielastic $e-d$ scattering is defined as,
$W^2 = M^2 + 2 M(E-E') -Q^2$, where $M$ is the nucleon mass; $W^2 = M^2 =0.88$
GeV$^2$ at the quasielastic peak.
The measured quasielastic $e-d$ cross-section per nucleon, 
$\sigma(E,E',\theta)$ converted to the reduced cross section is written as:

\begin{equation}
\sigma_{R} = \epsilon (1 + \tau') {\frac{\sigma(E,E',\theta)}{\sigma_{M}}} 
=R_{T} + \epsilon R_{L},
\end{equation}
where $\sigma_{M}$ is the Mott cross section defined earlier, 
and $R_T$ and $R_L$ are the 
transverse and longitudinal nuclear response functions, respectively.
In the plane wave impulse approximation (PWIA) the quasielastic 
$R_T$ response function 
is proportional to ${G^{n}_{M}}^2 + {G^{p}_{M}}^2$, and 
the $R_{L}$ response function is proportional
to ${G^{n}_{E}}^2 + {G^{p}_{E}}^2$.  Thus, the extraction of the 
neutron electromagnetic form factor
requires the separation of the $R_L$ and the $R_T$ response 
functions using the Rosenbluth technique, followed by the subtraction of 
the proton contribution in PWIA.

Until recently, most data on $G_M^n$ had been deduced from 
quasi-elastic electron-deuteron scattering. For inclusive
measurements~\cite{hughes,gross,esau,arnold,lung},
the procedure requires the separation of the longitudinal 
and transverse cross sections and the subsequent subtraction of a large
proton contribution. Thus, it suffers from large theoretical uncertainties
due in part to the deuteron model employed and in part 
to corrections for final-state interactions (FSI), MEC effects, and  
relativistic corrections. The proton subtraction can be avoided by 
measuring the neutron in coincidence $d(e,e'n)$, 
but the difficulty is associated with the absolute neutron detection 
efficiency calibration. This method was used by 
Stein {\it et al.}~\cite{stein}, Bartel {\it et al.}~\cite{bartel},
and most recently at MIT-Bates Laboratory
by Markowitz {\it et al.}~\cite{Mark93}. In addition,  
the anti-coincident $d(e,e'\bar{p})$ 
measurement, where the absence of the proton detection is required, was 
also carried out in the past~\cite{budnitz,hanson}. 

\begin{figure}[htbp]
\centerline{\epsfig{file=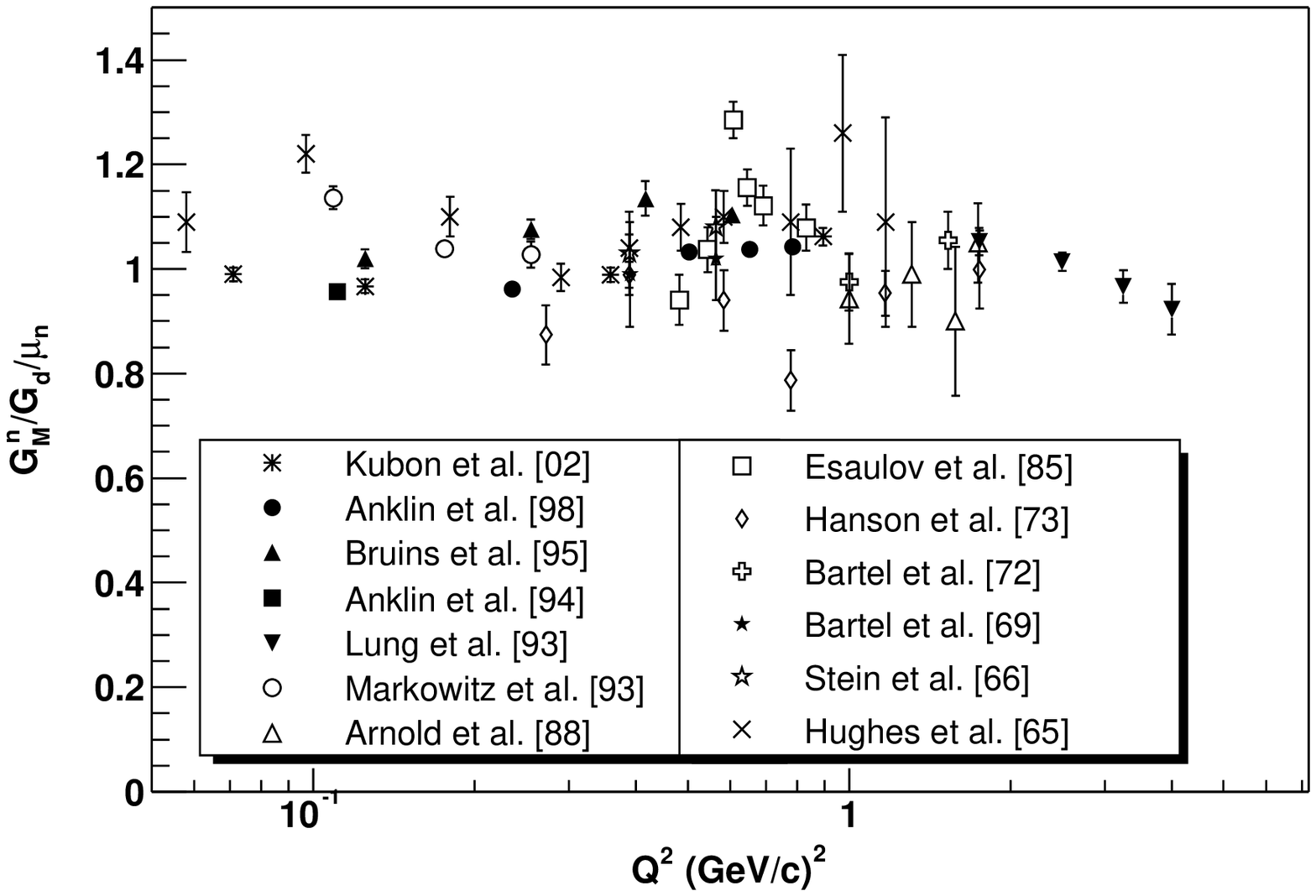,height=6.0cm,width=12.0cm}}
\fcaption{{\footnotesize\it  The published neutron magnetic form factor 
data from unpolarized deuterium experiments [23-36]. The data are plotted as 
$G^{n}_{M}/\mu_n$, in units of the standard dipole form factor 
$G_D$, as a function of $Q^{2}$.}}
\label{fig:gmndata}
\end{figure}

The sensitivity to nuclear structure can be greatly reduced by 
taking the cross-section ratio of 
$d(e,e'n)$ to $d(e,e'p)$ at quasi-elastic kinematics.  Several recent
experiments~\cite{Ankl94,Brui95,Ankl98,kub01} have employed this
technique to extract $G_M^n$ with uncertainties of $<$2\%~\cite{Ankl98,kub01}
for $Q^2$-values from 0.1 to 0.8 (GeV/c)$^2$. While such level of 
precision is excellent, there is considerable disagreement among the 
results~\cite{Mark93,Ankl94,Brui95,Ankl98,kub01} concerning the absolute 
value of $G_M^n$. All coincidence experiments 
require an absolute calibration of the neutron detection efficiency, which
is a challenging task. Most recently, the $G^n_M$ measurement using the 
ratio technique has been extended to a $Q^2$ value of 
above 4.0 (GeV/c)$^2$ at Jefferson Lab~\cite{brook2}.
While discrepancies among the deuterium experiments 
described above may be understood~\cite{jordan},
additional data on $G_M^n$, preferably obtained using a complementary
method, are highly desirable.
The inclusive quasi-elastic reaction $^3\vec{\rm He}(\vec{e},e')$ 
provides just such an alternative approach.  In 
comparison to deuterium experiments, this technique employs a different 
target and relies on polarization degrees of freedom, and is therefore 
subject to completely different systematics. We will discuss this type of 
double-polarization measurements in the next section. 
The world data on $G^n_M$ from unpolarized $ed$ quasielastic scattering
experiments are shown in Fig.~\ref{fig:gmndata}.

\vspace{0.7in}


\section{Double-Polarization Electron Scattering Experiments}

Polarization degrees of freedom in electron scattering have proven to
be very useful in extracting information about small amplitude
processes by isolating terms sensitive to the interference of the
small amplitude with a much larger amplitude. The ability to
selectively isolate certain combinations of amplitudes that either polarized
electron beams or polarized targets (recoil polarimeters) 
have historically provided when
used in isolation, is significantly enhanced by using them together.
Thus, polarized electron scattering from polarized targets, and
polarization transfer measurements using recoil polarimeters can
provide important information on the nucleon spin structure and the
electromagnetic structure of the nucleon. 

The development of polarized beams, polarized targets and recoil polarimeters
has allowed more complete studies of electromagnetic structure
than has been possible
with unpolarized reactions. In quasielastic scattering, the spin degrees
of freedom introduce new response functions into the differential 
cross section, which provides additional information 
on nuclear structure~\cite{donnelly}. Such studies are especially important for
probing the neutron electromagnetic structure because of the lack of free
neutron targets in nature. Section 4.1 contains a brief 
overview of experimental techniques employed in double 
polarization experiments. 

\subsection{Experimental Techniques in Double-Polarization Experiments}

\subsubsection{Polarized electron beam} 

Polarized electrons can be produced by various techniques~\cite{pol_electron}. 
Photoemission from GaAs has several advantages: high peak current,
helicity reversal by optical means, and the feasibility of building a 
GaAs source with small transverse phase space or emittance, an important 
requirement for some electron accelerator machines. 

The GaAs polarized electron source works through photo-emission of electrons
that have been polarized through optical pumping. 
The electron beam is generated by illuminating a GaAs cathode, 
which is placed in ultra-high vacuum, with high-intensity circularly 
polarized laser light. Strained GaAs cathodes are frequently used nowadays 
and have advantage over conventional bulk GaAs cathodes.
The strain creates a gap in the different sub-levels of the 
$P_{3/2}$ electrons in the valence bands of the GaAs~\cite{Prepost95}. 
Left circularly polarized light with the right frequency, 
incident on the GaAs, only drives transition between the $P_{3/2}$ m=3/2 
state in the valence band and the $S_{1/2}$ m= 1/2 level of 
the conduction band. From there the polarized electrons 
diffuse to the surface and escape into the surrounding vacuum. 
Typically, the surface of the GaAs cathode is treated with cesium 
to create a negative electron affinity. The polarization of source electrons 
is consequently as high as $80\%$. 

The electrons from the polarized source are longitudinally polarized, 
and are then accelerated in the linac.
The electron beam polarization can be
determined either by using a M\mbox{\o}ller polarimeter which measures the 
asymmetry from polarized electrons scattering off polarized atomic 
electrons in a magnetized iron foil, or by using a 
Compton polarimeter in which the asymmetry from circularly polarized 
laser light backscattering from the polarized electron beam is measured. 
In some of the high precision asymmetry measurements, 
both polarimeters might be employed to minimize systematic 
uncertainties in the
determination of the electron beam polarization.

\subsubsection{Polarized targets}

Polarized targets for studies of nucleon structure can be categorized as 
polarized proton targets and polarized effective neutron targets 
because of the lack of free, stable neutrons in nature. 
For the purpose of a brief 
overview of polarized targets, the discussion below is organized 
according to gaseous and solid targets. 

There are two general types of polarized
gaseous targets: external and internal. External targets are sealed 
targets with windows. They are much more dense than internal targets
in order to achieve a practical luminosity 
with relatively low incident electron beam currents. The windowless 
internal targets placed in an electron storage ring with very 
high electron beam currents are open flowing systems, thus with 
much thinner average target thicknesses compared to external targets.
 
The internal gas targets are typically polarized by either 
optical pumping or by the atomic beam source method.
Optical pumping is widely used to polarize a sample of atoms by 
transferring angular momentum from a pump light beam, typically a laser
beam, to the sample atoms.
The atomic beam source (ABS) method is based on the well-known principle of
Stern-Gerlach separation and radio frequency (RF) transitions.
One can employ either the atomic beam source method which is a
well-established technique, or the optical pumping method
for polarized hydrogen (proton) and deuterium (effective neutron) internal
targets.  The typical
atomic flux an ABS can feed a storage target cell is $\sim 6\times 10^{16}$
atoms/second and the proton and deuteron nuclear vector polarization can be
as high as $90\%$. 

An internal polarized H/D gas target is also feasible with
spin-exchange optical pumping method. Such a target works 
in the following way: 
polarized laser light optically pumps alkali atoms.
Hydrogen (deuterium) atoms are polarized through spin-exchange
collisions with the alkali atoms.
At large hydrogen (deuterium) densities, frequent H-H (D-D) collisions
enhance the probability of hyperfine interactions, and the
system approaches spin temperature equilibrium~\cite{ART:WAA93}.
The nucleus is polarized in spin temperature equilibrium,
in which hydrogen nuclear and atomic polarizations
are the same and, in the case of deuterium nuclear vector polarization 
exceeds the atomic polarization.
Spin-exchange optical pumping of alkali-metal atoms 
with hydrogen (deuterium) is an efficient way to polarize
hydrogen (deuterium) atoms and atomic fluxes as high as 
$10^{18}$ atoms/second are typical with this type of source, though with much
lower nuclear polarization than that of ABS, 
a $50\%$ nuclear vector polarization  
for hydrogen and $60\%$ for deuterium can be anticipated.

Polarized $^3$He is useful for studying the neutron
structure because its ground state is dominated by a spatially symmetric $S$
wave in which the proton spins cancel and 
the spin of the $^3$He
nucleus is predominantly carried by the unpaired neutron alone.
Internal polarized $^3$He gas targets can be produced in principle 
either by using optical
pumping methods or atomic beam source method. 
However, only metastability-exchange optical pumping 
has been used up to now in producing internal targets of practical 
thicknesses for scattering experiments. 
In the case of $^{3}$He, direct optical 
pumping between its ground state and the first excited state is not possible 
because of the energy difference involved. 
The metastability-exchange optical pumping
involves optical pumping of 2$^{3}S_1$ metastable state 
atoms, then transferring the polarization to $^3$He ground state atoms
through metastability-exchange collisions.

The only external gaseous targets that exist for scattering experiments are 
polarized $^3$He targets based on both the metastability-exchange optical 
pumping technique and the spin-exchange optical pumping technique in which 
rubidium vapor is typically used as the spin-exchange medium. While 
spin-exchange optical pumping is capable of producing high density targets, 
metastability-exchange optical pumping combined with mechanical
compression produce polarized $^3$He targets of comparable densities.
External polarized $^3$He targets based on these techniques have been used 
successfully at electron accelerator laboratories around the world. For more
details about these gaseous polarized targets, we refer the reader 
to a review article by
Chupp, Holt and Milner~\cite{chm94}.

Polarized solid targets have been used widely in lepton scattering 
experiments probing nucleon structure, particularly in deep-in-elastic 
lepton scattering experiments studying nucleon spin structure. The SMC 
experiment~\cite{smc} at CERN used a low intensity muon 
beam incident on a very thick 
target of butanol or deuterated butanol, which was cooled by a 
powerful dilution refrigerator. 
Polarization is achieved by the technique of
dynamic nuclear polarization, which works in the following way.
A hydrogeneous compound is doped with a small concentration ($10^{-4}$) of 
unsaturated electron spins. In a large magnetic field and low temperatures
these electrons are polarized from the Boltzmann distribution. At a temperature
of 1 K and a magnetic field of 5 Tesla, the electron polarization is
$99.8\%$. The relaxation time of the electrons in the higher-energy state
is short ($\sim 10^{-3}$ second) due to the interaction 
of the electrons with the 
lattice. The electrons therefore flip back and are available for further
spin-flip transitions with other protons (deuterons). The relaxation time of
the nuclei, on the other hand, is long ($\sim 10^3$ second). 
This allows the nuclei
to accumulate preferentially in a state selected by the frequency of the
microwaves, thus leading to a high polarization of the sample.
Experiments at SLAC (E143, E155)~\cite{e143,e155} and 
Jefferson Lab used polarized solid 
$NH_3$ and $ND_{3}$ targets~\cite{clas,zhu} based on the same technique of
dynamic nuclear polarization. We refer interested readers 
to Ref.~\cite{crabb-day} for details 
on these targets.

\subsubsection{Recoil polarimeters}

Focal plane polarimeters were standard equipment at intermediate energy 
proton accelerators, such as LAMPF~\cite{lampf}, TRIUMF~\cite{triumf}, 
SATURNE~\cite{saturne}, and PSI~\cite{psi}. 
Sensitivities of spin observables to small amplitudes were demonstrated
by experiments carried out at these facilities using focal plane 
polarimeters. In recent years, proton
recoil polarimeters have been employed in a number of experiments at electron
accelerator facilities, such as the MIT-Bates Laboratory~\cite{batesfpp}, 
the Mainz laboratory~\cite{dieterich} and the Jefferson 
Laboratory~\cite{mjones,gayou1,gayou}. 

Proton polarimeters are based on nuclear scattering from an analyzer 
material like carbon. The proton-nucleus spin-orbit interaction results in an
azimuthal asymmetry in the scattering distribution which can be analyzed to
obtain the proton polarization and spin orientation. 
A typical proton recoil polarimeter consists of two front detectors to
track incident protons, followed by a carbon analyzer and two rear detectors
to track scattered particles.
Recoil neutron polarimeters which have been used in experiments to extract
neutron electric form factor are based on the same physics principle and are
constructed in similar ways.

\vspace{0.7in}
\begin{figure}[htbp]
\centerline{\epsfig{file=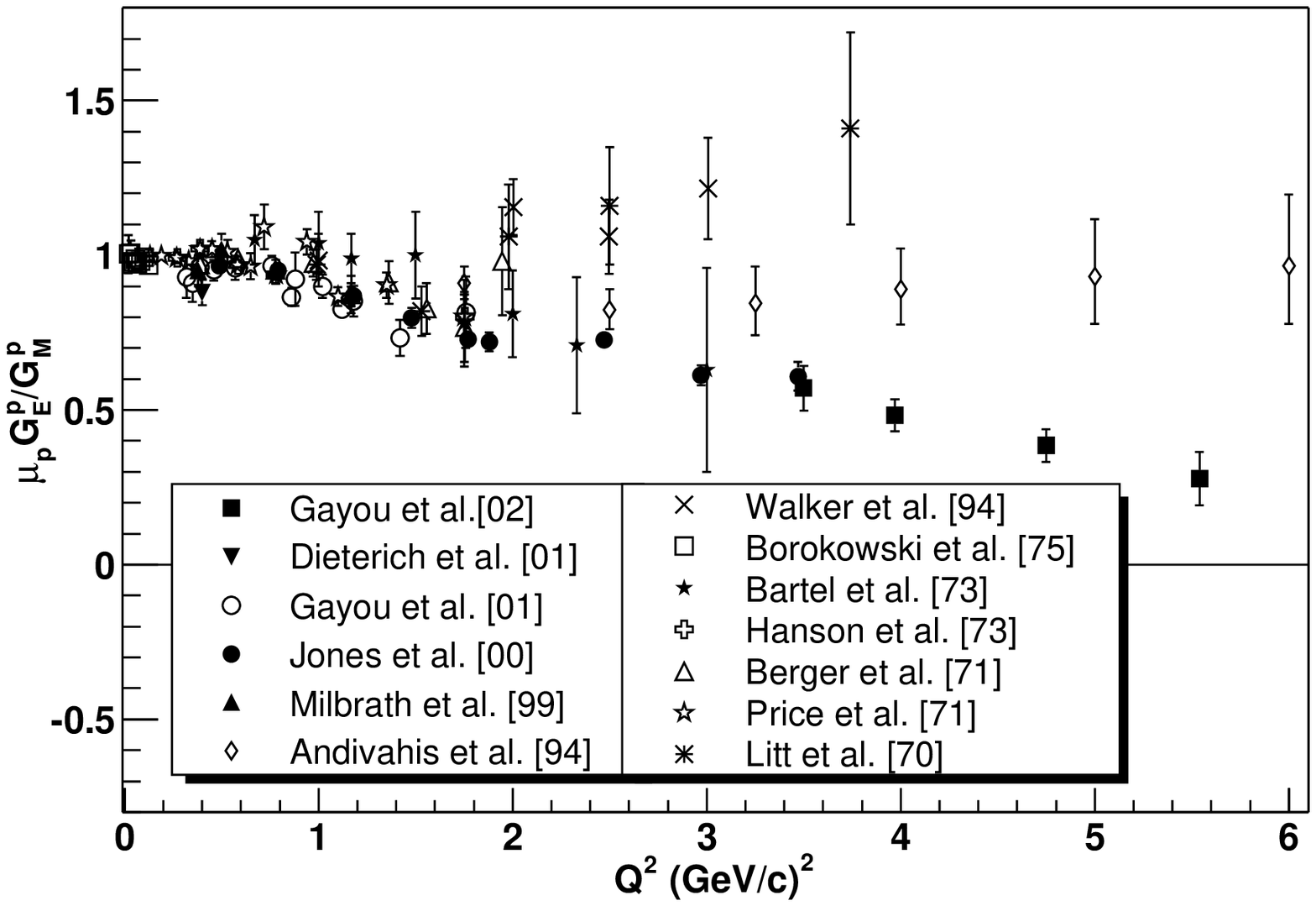,height=6.0cm,width=12.0cm}}
\fcaption{{\footnotesize\it World data on the proton electric and magnetic 
form factor ratio, ${\frac{\mu G^p_E}{G^p_M}}$,
as a function of $Q^2$. 
The high precision data from polarization transfer 
measurements at Jefferson Lab by Gayou {\it et al.}~\cite{gayou} 
and by Jones {\it et al.}~\cite{mjones} plotted with 
statistical uncertainties only, show different $Q^2$-dependence 
of the form factor 
ratio from those obtained by the Rosenbluth method 
from unpolarized cross-section measurements, at large values of $Q^2$.}}
\label{fig:fpp}
\end{figure}

\subsection{Polarized Electron-Proton Elastic Scattering}

While precise information on $G^p_E$ and $G^p_M$ is important for 
understanding the underlying electromagnetic structure of the proton, it is
also very interesting to study the ratio of these two form factors, 
${\frac{\mu_{p}G^p_E}{G^p_M}}$ as a function of $Q^2$.
Any $Q^2$ dependence in the proton form factor ratio would suggest 
different charge and magnetization spatial distributions inside the proton.
Double polarization experiments allow precise measurement of the 
proton form factor
ratio directly. Furthermore, by combining 
polarization measurements with differential cross-section measurements, 
one can determine $G^p_E$ and $G^p_M$ separately with significantly 
improved precision compared to the conventional Rosenbluth separation 
technique.

In the one-photon-exchange Born approximation, 
the elastic scattering of longitudinally 
polarized electrons from unpolarized protons 
results in a transfer of polarization to the recoil 
proton with only two nonzero components: $P_t$ perpendicular to, 
and $P_l$ parallel to the proton momentum in the scattering 
plane~\cite{arnold2}. 
The form factor ratio can be
determined from a simultaneous measurement of the two recoil polarization
components in the scattering plane:
\begin{equation}
{\frac{G^p_E}{G^p_M}} = - {\frac{P_{t}}{P_{l}}} {\frac{E+E'}{2M}} 
\tan({\frac{\theta}{2}}),
\end{equation}

where $E$ and $E'$ are the incident and scattered electron energy, 
respectively, and $\theta$ is the electron scattering angle.

Although no dramatic $Q^2$-dependence in the proton form factor 
ratio has been observed from unpolarized 
measurements, new data from a polarization transfer experiment~\cite{mjones}, 
which measured this ratio directly shows very intriguing behavior at higher
$Q^2$, i.e., $G^p_E$ falls off much faster than $G^p_M$ as a function 
of $Q^2$. The intriguing result on ${\frac{\mu_{p} G^p_E}{G^p_M}}$ at 
high $Q^2$ values from Jefferson Lab~\cite{mjones} has
motivated much interest on this subject both experimentally~\cite{gayou1} 
and theoretically.
A more recent polarization transfer experiment at Jefferson Lab~\cite{gayou} 
showed that the downward trend continues out to $Q^2=$ 5.6 (GeV/c)$^2$.
Prior to the Jefferson Lab experiments, the recoil polarization technique
was employed at the MIT-Bates laboratory~\cite{batesfpp}; 
and it has also been used more recently at Mainz~\cite{dieterich}. 
Fig.~\ref{fig:fpp} shows the world data on the proton 
form factor ratio with statistical uncertainties only 
from these recoil polarization measurements, 
as well as those obtained from unpolarized experiments.
The new data from polarization experiments at Jefferson Lab show unprecedented 
precision compared to the data from unpolarized experiments. These new 
data demonstrated an intriguing $Q^2$ dependence, dramatically 
different from the behavior observed in the unpolarized 
Rosenbluth separation experiments. Recently, a new experiment was completed 
at Jefferson Lab~\cite{segal} aiming at resolving such a discrepancy by using 
the so-called super Rosenbluth separation technique. Results from this 
experiment are anticipated in the near future.

\begin{figure}[htbp]
\centerline{\epsfig{file=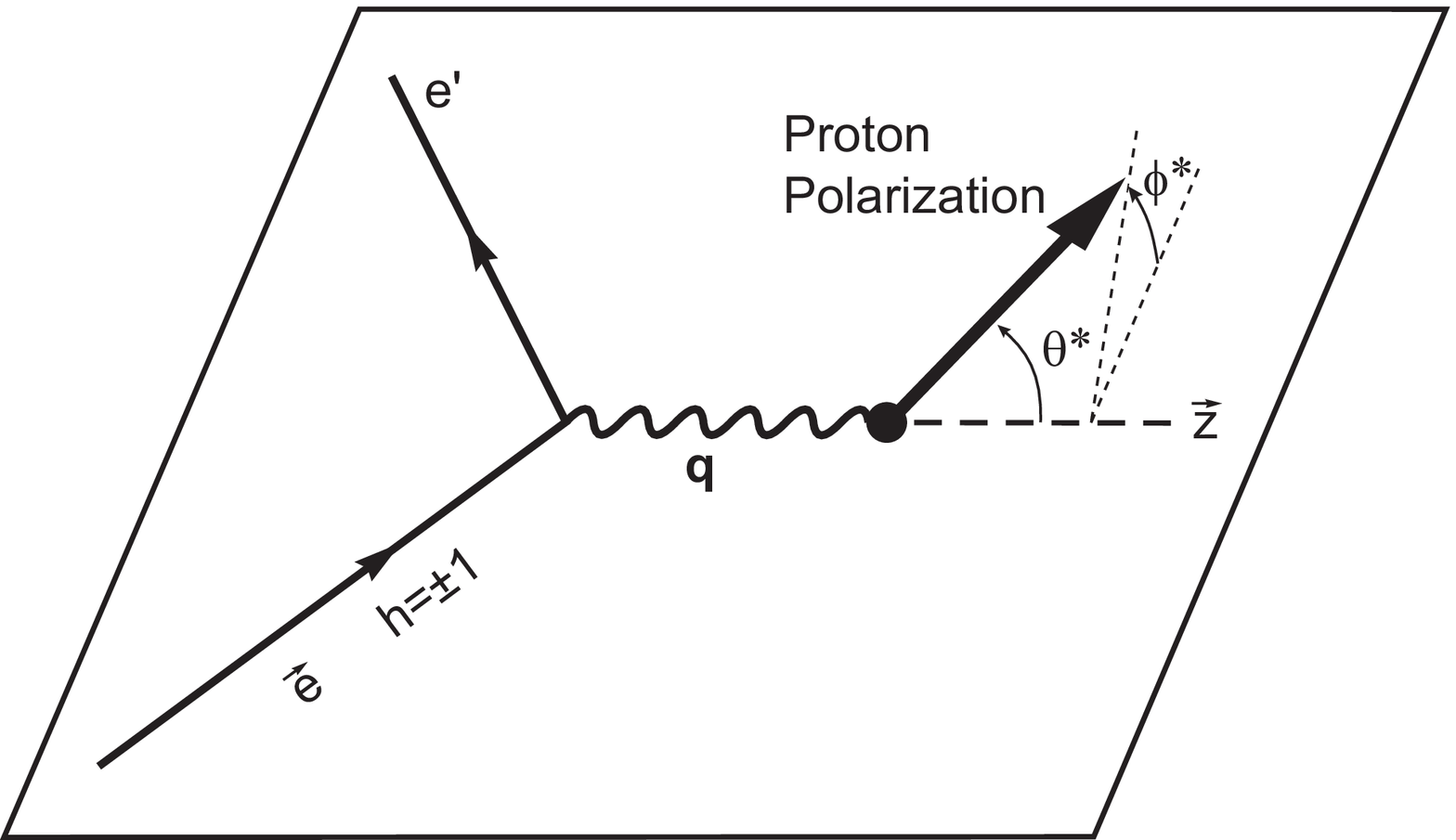,height=5.0cm,width=10.0cm}}
\fcaption{{\footnotesize\it The one-photon-exchange diagram for spin-dependent
electron-nucleon elastic scattering.}}
\label{spin_dependent}
\end{figure}

Alternatively, one can determine the proton form factor ratio by 
measuring the asymmetry of longitudinally polarized electrons scattered from 
a polarized proton target. The one-photon-exchange diagram for such 
spin-dependent elastic scattering is shown in Fig.~\ref{spin_dependent}.
For longitudinally polarized electrons scattering from a polarized proton 
target, the differential cross section can be written as~\cite{donnelly}:

\begin{equation}
{\frac{d\sigma}{d\Omega}} = \Sigma + h  \Delta \>,
\end{equation}

where $\Sigma$ is the unpolarized differential cross section 
given by Eqn.~\ref{eq:unpol}, $h$
is the electron helicity, and $\Delta$ is the 
spin-dependent differential cross section given by:
\begin{equation}
\Delta= - \sigma_{Mott} f^{-1}_{recoil} [ 2 \tau v_{T'} 
\cos \theta^{*} {G^{p}_{M}}^{2} - 2 \sqrt{2 \tau (1 + \tau)} v_{TL'}
\sin\theta^{*} \cos\phi^{*} G^{p}_{M} G^{p}_{E}]\>,
\end{equation}

\vspace{0.5in}

where $\theta^{*}$ and $\phi^{*}$ are the polar and azimuthal proton
spin angles defined with respect to the three-momentum transfer vector
$\vec{q}$ and the scattering plane 
(Fig.~\ref{spin_dependent}), and 
$v_{T'}$ and $v_{TL'}$ are kinematic factors~\cite{donnelly}.

The spin-dependent asymmetry $A$ is defined as:
\begin{equation}
    A = {\frac{\sigma^{h+}-\sigma^{h-}}{\sigma^{h+}+\sigma^{h-}}},
\end{equation}

where $\sigma^{h{\pm}}$ denotes the differential cross section for the two 
different helicities of the polarized electron beam.
The spin-dependent asymmetry $A$ can be written 
in terms of the polarized
and unpolarized differential cross-sections as: 
\begin{equation}
A = {\frac{\Delta}{\Sigma}} = - {\frac{ 2\tau v_{T'} \cos\theta^{*}
{G^{p}_{M}}^2 - 2 \sqrt{2 \tau ( 1 + \tau)} v_{TL'} \sin\theta^{*}
\cos\phi^{*} G^{p}_{M} G^{p}_{E}}{(1 + \tau) v_{L} {G^{p}_{E}}^{2}  + 2
\tau v_{T} {G^{p}_{M}}^{2}}} \>.
\label{eqn:asymmetry}
\end{equation}
The experimental asymmetry $A_{exp}$, is related to the spin-dependent
asymmetry of Eqn.~\ref{eqn:asymmetry} by the relation 
\begin{equation}
A_{exp} = P_{b} P_{t} A \>,
\end{equation}
where $P_{b}$ and $P_{t}$ are the beam and target polarizations,
respectively. A determination of the ratio 
$\frac{G^{p}_{E}}{G^{p}_{M}}$, independent of the knowledge
of the beam and target polarization, 
can be precisely obtained by forming the so-called super ratio 

\begin{equation}
R = {\frac{A_1}{A_2}} = {\frac{ 2\tau v_{T'} \cos\theta_{1}^{*}
{G^{p}_{M}}^2 - 2 \sqrt{2 \tau ( 1 + \tau)} v_{TL'} \sin\theta_{1}^{*}
\cos\phi_{1}^{*} G^{p}_{M} G^{p}_{E}} {2\tau v_{T'} \cos\theta_{2}^{*}
{G^{p}_{M}}^{2} - 2 \sqrt{2 \tau ( 1 + \tau)} v_{TL'}
\sin\theta_{2}^{*} \cos\phi_{2}^{*} G^{p}_{M} G^{p}_{E}}} \>, 
\end{equation}
where $A_1$ and $A_2$ are elastic electron-proton scattering asymmetries
measured at the same $Q^2$ value, but two different proton 
spin orientations corresponding to $(\theta_{1}^{*}, \phi_{1}^{*})$ and
$(\theta_{2}^{*}, \phi_{2}^{*})$, respectively. 
For a detector configuration that is symmetric with respect to the incident 
electron momentum direction,
$A_{1}$ and $A_{2}$ can be measured simultaneously by forming
two independent asymmetries with respect to either
the electron beam helicity or the target spin orientation 
in the beam-left and beam-right sector of the detector system, respectively.
Thus, the proton form factor ratio can be determined with high precision using 
this technique, and an experiment is currently planned at the MIT-Bates
Laboratory~\cite{blast}.  Such an experiment is essential because it 
employs a different experimental technique that has different 
systematic uncertainties than those from recoil proton polarization 
experiments, and those from Rosenbluth and super Rosenbluth measurements.

\subsection{Polarized Quasielastic Electron-Nucleus Scattering and Neutron
Electromagnetic Form Factors}

\subsubsection{Experiments with Deuterium Targets}

Measurements of the neutron electric form factor are
extremely challenging because of the lack of free neutron targets, 
the smallness of the $G^n_E$, and the dominance of the magnetic contribution
to the unpolarized differential cross-section. A promising approach to measure
$G^n_E$ is by using polarization degrees of freedom. 
For coincidence elastic scattering of longitudinally polarized electrons 
from ``free'' neutrons, $n(\vec{e},e'\vec{n})$ process, 
the recoil neutron polarization is given by~\cite{arnold2}: 

\begin{equation}
P_{z} = {\frac{hP_e}{\epsilon {G^n_E}^2 + {\tau G^n_M}^2}}
{\frac{E+E'}{M}}\epsilon\sqrt{\tau(1+\tau)}{G^n_M}^2
\tan^{2}({\frac{1}{2}}\theta)
\end{equation}

\begin{equation}
P_{x} = {\frac{hP_e}{\epsilon {G^n_E}^2 + {\tau G^n_M}^2}}
(-2 \epsilon \sqrt{\tau(1+\tau)}{G^n_EG^n_M})\tan({\frac{1}{2}}\theta),
\end{equation}
where $P_{z}$ and $P_{x}$ are the recoil neutron polarization component along
the recoil neutron momentum direction, and transverse to it, respectively, in
the scattering plane. Here $P_e$ denotes the absolute value of the electron 
polarization and $h=\pm 1$ denotes the electron helicity.
In the case of a ``free'' neutron target, the normal component $P_y$ 
is zero due to time and parity invariance. However, 
$P_y$ is typically not zero because an effective neutron (nuclear) target 
must be employed and final state interactions play a role.
Thus, by forming the ratio of ${\frac{P_x}{P_z}}$, the unpolarized 
cross-section as well as the electron polarization cancels out, providing 
sensitive experimental access to the neutron form factor ratio,
\begin{equation}
{\frac{P_x}{P_z}} = {\frac{-2M}{E+E'}}{\left[\tan({\frac{1}{2}}\theta)\right]}^{-1}
{\frac{G^n_E}{G^n_M}}.
\label{eqn:genpol}
\end{equation}

Eqn.~\ref{eqn:genpol} is only valid in the PWIA picture and corrections need
to be applied when effective neutron targets are used. 
The deuteron is the simplest effective neutron target and it is also
the most effective neutron target for recoil
neutron polarization measurements. Correcting for nuclear binding effect, 
relativity, reaction mechanisms such as FSI and MEC effects is essential
in order to extract reliable information on $G^n_E$ from quasielastic 
$d(\vec{e},e'\vec{n})$ reaction. Furthermore, precise information 
on $G^n_M$ is crucial in order to extract $G^n_E$ because it is the ratio 
${\frac{G^n_E}{G^n_M}}$ to first order 
that is determined from such a recoil neutron 
polarization measurement. 
Experiments with longitudinally polarized electron beams and recoil 
neutron polarimeters have been carried out at MIT-Bates~\cite{eden} 
and Mainz~\cite{ostrick,herberg} in the relatively low $Q^2$ region,
and $G^{n}_{E}$ has been extracted from the $d(\vec{e},e'\vec{n})$ process,
using the state-of-the-art two-body calculations 
by Arenh\"{o}vel~\cite{arenhovel}.
Most recently, such an approach has been employed at Jefferson Lab up 
to a $Q^2$ value of 1.5 (GeV/c)$^2$~\cite{madey}.

Alternatively, one can employ a polarized deuteron target to
probe the neutron electric form factor by the $\vec{d}(\vec{e},en)$
reaction. The scattering cross-section for longitudinally polarized 
electrons from a polarized deuteron target for the
$\vec{d}(\vec{e},en)$ reaction can be written as~\cite{arenhovel}:
\begin{equation}
S=S_{0} \left\{ 1+P^{d}_{1}A^{V}_{d}+ P^{d}_{2}A^{T}_{d} 
+ h(A_{e} + P^{d}_{1}A^{V}_{ed} + P^{d}_{2}A^{V}_{ed})\right\},
\end{equation}
where $S_{0}$ is the unpolarized differential cross section, $h$ the
polarization of the electrons, and $P^{d}_{1}$ ($P^{d}_{2}$) the
vector (tensor) polarization of the deuteron. $A_e$ is the beam
analyzing power, $A^{V/T}_d$ the vector and tensor analyzing powers,
and $A^{V/T}_{ed}$ the vector and tensor spin-correlation parameters.
The polarization direction of the deuteron is defined with respect to the
three-momentum transfer vector, $\vec{q}$. The vector spin-correlation 
parameter $A^{V}_{ed}$ contains a term representing the interference 
between the small
neutron electric form factor and the dominant neutron magnetic form
factor, when the target spin is perpendicular to
the $\vec{q}$ vector direction. Thus, the spin-dependent asymmetry
(defined with respect to the electron beam helicity)
from the $\vec{d}(\vec{e},en)$ reaction for vector polarized deuteron
gives access to the quantity ${\frac{G^n_E}{G^n_M}}$ to first order 
when the target spin direction is aligned perpendicular 
to $\vec{q}$. Such experiments
are extremely challenging since they involve both neutron detection and a
vector polarized deuteron target.
 
Recently, the neutron electric form factor was extracted for the first 
time~\cite{nikhef} from a $\vec{d}(\vec{e},e'n)$ measurement in which a 
vector polarized deuteron target from an atomic beam source was employed.
Most recently, a $\vec{d}(\vec{e},e'n)$ experiment~\cite{zhu}
using a dynamically polarized solid deuterated ammonia target was carried out
at Jefferson Lab and $G^n_E$ was extracted at a $Q^2$ value of 0.5 (GeV/c)$^2$.
The precision of $G^n_E$ from these polarization experiments is significantly
better than those by Platchkov {\it et al.}~\cite{platchkov} 
from the unpolarized electron-deuteron elastic scattering measurement.
Thus, by using polarization degrees of freedom and coincidence
detections of scattered electrons and recoil neutrons, 
the dominant neutron magnetic contribution and
the proton contribution to the scattering process 
are suppressed and more precise information on the neutron
electric form factor can be extracted. 

\subsubsection{Quasielastic $^{3}\vec{He}(\vec{e},e')$ and 
$^{3}\vec{He}(\vec{e},e'n)$ Reactions}

\begin{figure}[htbp]
\centerline{\epsfig{file=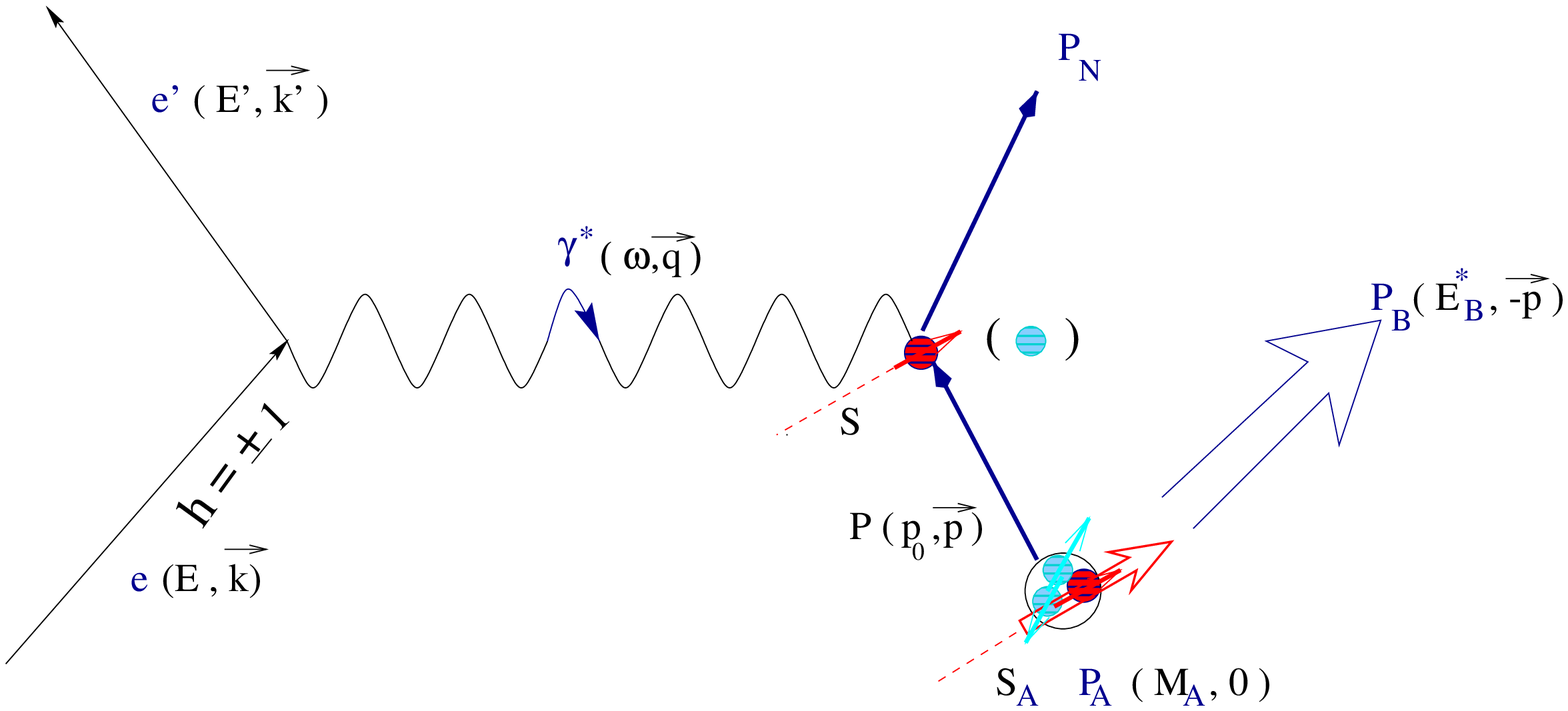,height=6.0cm,width=12.0cm}}
\fcaption{{\footnotesize\it The one-photon-exchange diagram for spin-dependent
quasielastic electron-$^3$He scattering in the plane-wave impulse 
approximation picture.}}
\label{he3pol}
\end{figure}

A polarized $^3$He nucleus is effective in studying the neutron
electromagnetic form factors because of the unique spin structure of
the $^3$He ground state, which is dominated by a spatially symmetric $S$
wave in which the proton spins cancel and the spin of the $^3$He
nucleus is carried by the unpaired neutron~\cite{BW84,frier90}. 
Fig.~\ref{he3pol} shows the one-photon exchange diagram for 
longitudinally polarized electrons scattering off from a polarized $^3$He 
nuclear target at the quasielastic kinematics in the PWIA picture. 
For inclusive measurement, only the scattered electron is detected; 
for the coincidence $^{3}\vec{He}(\vec{e},e'n)$ 
measurement, both the scattered electron and the 
knockout neutron are detected.

The spin-dependent contribution to
the inclusive $^3\vec{\rm He}(\vec{e},e')$ cross section 
is completely contained in 
two spin-dependent nuclear response functions, 
a transverse response $R_{T'}$ and a
longitudinal-transverse response $R_{TL'}$~\cite{donnelly}.
These appear in addition to the
spin-independent longitudinal and transverse responses $R_{L}$ and
$R_{T}$. These spin-dependent response functions $R_{T'}$ and $R_{TL'}$ 
can be isolated experimentally by forming the spin-dependent 
asymmetry $A$ defined previously with respect to the electron beam helicity.
In terms of the nuclear response functions, $A$ can be 
written~\cite{donnelly}: 
\begin{equation}
\label{asym}
A = \frac{-(\cos{\theta^{*}}\nu_{T'}R_{T'} +
  2\sin{\theta^{*}}\cos{\phi^{*}}\nu_{TL'}R_{TL'})}{\nu_{L}R_{L} +
  \nu_{T}R_{T}}
\end{equation}  
where the $\nu_{k}$ are kinematic factors, and $\theta^{*}$ and
$\phi^{*}$ are the target spin angles defined previously. 
The response functions
$R_{k}$ depend on $Q^{2}$ and the electron energy 
transfer $\omega$. 
By choosing $\theta^\ast = 0$, {\it i.e.} by orienting the target
spin parallel to the momentum transfer ${\vec q}$, one selects the
transverse asymmetry $A_{T'}$ (proportional to $R_{T'}$); 
by orienting the target spin perpendicular to the momentum 
transfer ${\vec q}$ ($\theta^\ast = 90$, $\phi^\ast=0$), one selects the
transverse-longitudinal asymmetry $A_{TL'}$ (proportional to $R_{TL'}$).

\vspace{0.in}
\begin{figure}[htbp]
\centerline{\epsfig{file=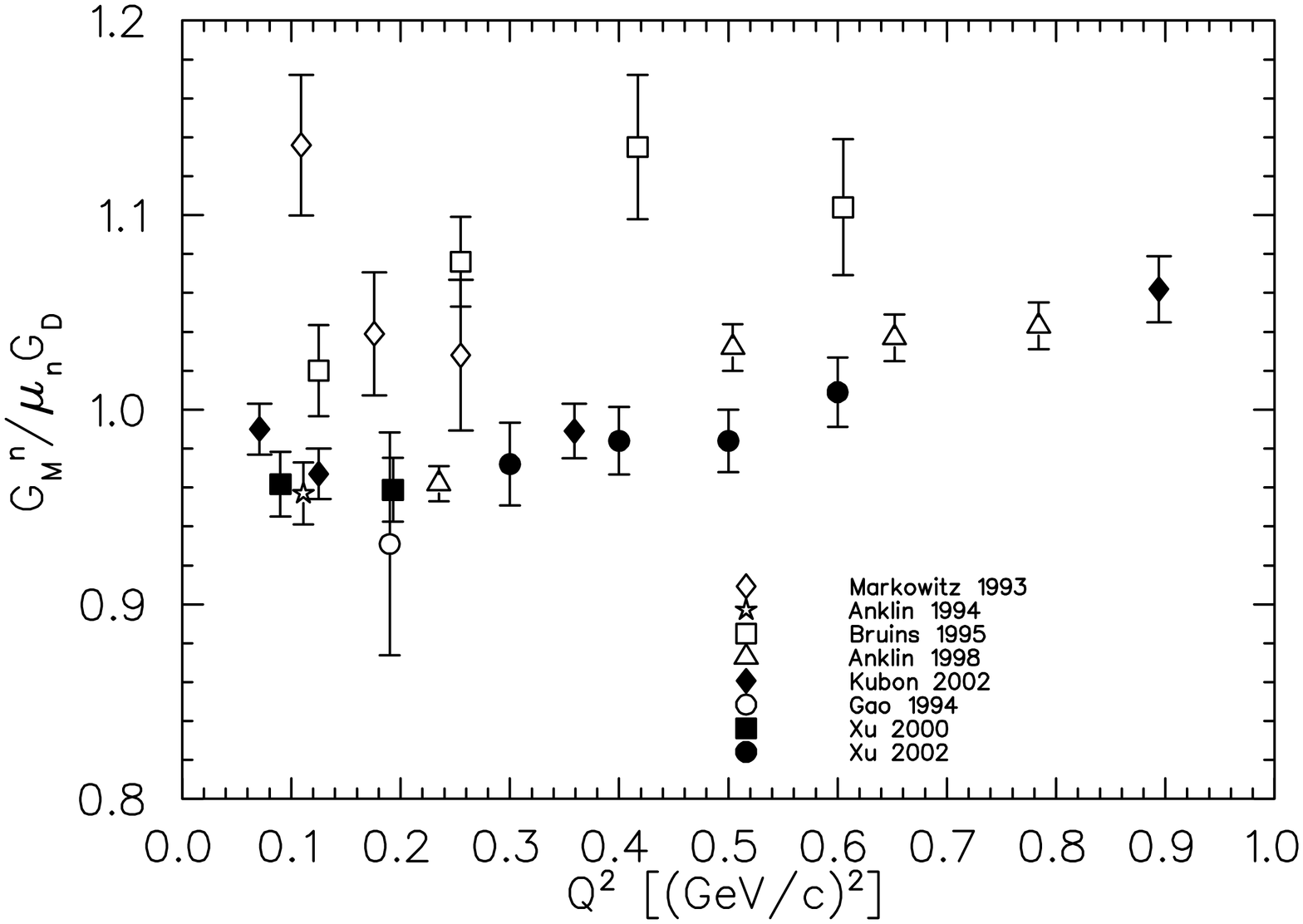,angle=90,
height=7.0cm,width=11.0cm}}
\fcaption{{\footnotesize\it The neutron magnetic form factor $G_{M}^{n}$ data  
published since 1990, in units of the standard dipole form 
factor parameterization $G_D$, as a function of
$Q^{2}$. The data by Gao {\it et al.}~\cite{Gao94}, 
Xu {\it et al.}~\cite{xu,e95001_prc} 
are from experiments employing polarized $^3$He targets discussed in the 
text, plotted with statistical uncertainties only.
The rest are from experiments involving deuterium targets.
The Q$^2$ points of Anklin 94~\cite{Ankl94} and Gao 94~\cite{Gao94} 
have been shifted slightly for clarity.}}
\label{fig:gmn}
\end{figure}

Because the $^3$He nuclear spin is carried mainly by the neutron,
$R_{T'}$ at quasi-elastic kinematics contains a dominant neutron
contribution and is essentially proportional to
${(G_M^n)}^2$, similar to elastic scattering from a free neutron.
Unlike the free neutron, however, the unpolarized part of the
cross section (the denominator in Eq.\ (\ref{asym})) contains
contributions from both the protons and the neutron in the $^3$He nucleus at
quasielastic kinematics.
Therefore, $A_{T'}$ is expected to first order to have the form
${(G_M^n)}^2/(a+b{(G_M^n)}^2)$ in PWIA, where $a$ is much larger than
$b{(G_M^n)}^2$ in the low $Q^2$ region.  While
measurements of $G_M^n$ using deuterium targets enhance the
sensitivity to the neutron form factor by detecting the neutron in
coincidence, a similar enhancement occurs in inclusive scattering from
polarized $^3$He because of the cancellation of the proton spins in
the ground state. This picture has been confirmed by several 
PWIA calculations~\cite{ciofi,salme,rws93}, a more recent and
more advanced calculation which fully includes
FSI~\cite{ishi98}, and most recently by Golak {\it et al.}~\cite{golak4}, 
in which both the FSI and MEC effects have been taken into account.

The first experiments~\cite{CEJ90,AKT92} which investigated 
the feasibility of using polarized $^{3}$He targets to study the neutron 
electromagnetic structure were performed at the MIT-Bates Linear 
Accelerator Center.
Following these two experiments, the first measurement of $G^n_{E}$ from 
$^{3}\vec{\rm He}(\vec{e},e'n)$ was reported by 
Meyerhoff {\it et al.}~\cite{Meyerhoff}, and the neutron magnetic form factor 
at low $Q^{2}$ was
extracted for the first time~\cite{Gao94} from inclusive
measurement of the quasielastic transverse asymmetry, $A_{T'}$.
Recently, a high precision measurement of $A_{T'}$~\cite{xu} was carried out 
in Hall A at Jefferson Lab at six quasielastic kinematic settings 
corresponding to central $Q^2$ values of 0.1 to 0.6 (GeV/c)$^2$. 
A state-of-the-art non-relativistic Faddeev calculation~\cite{golak4} 
was employed in the extraction 
of $G^n_M$ at the two lowest $Q^2$ values of the experiment 
and a PWIA calculation~\cite{salme} 
was used in the extraction of $G^n_M$ for the remaining $Q^2$ values 
of the experiment~\cite{e95001_prc}.

The Faddeev calculation 
treats the $^3$He target state and the 3N scattering states   
in the nuclear matrix element 
in a consistent way by solving the corresponding 3N Faddeev 
equations~\cite{golak2}. 
The MEC effects were calculated using the prescription of 
Riska~\cite{riska85}, which includes $\pi$- and $\rho$-like exchange terms.
The Faddeev based formalism has been applied to other reaction
channels and good agreement has been found with experimental 
results~\cite{golak3}, in particular with the 
recent NIKHEF data on $A^{0}_{y}$ at 
$Q^2 = 0.16$ (GeV/c)$^2$ from the
quasielastic $^{3}\vec{\rm {He}}(\vec{e},e'n)$ process \cite{hans}, and
with the high precision $^3$He quasielastic asymmetry data in the 
breakup region~\cite{xiong} at low $Q^2$ 
from Jefferson Lab. 
This calculation, while very 
accurate at low $Q^2$, is not expected to be sufficiently precise 
for a reliable extraction of $G_M^n$ from the $^3$He asymmetry data
in a relatively higher $Q^2$ region because of its fully 
non-relativistic nature.

Fig.~\ref{fig:gmn} shows data from the Jefferson Lab 
experiment~\cite{xu,e95001_prc}, the 
earlier MIT-Bates experiment~\cite{Gao94}, and data from deuterium experiments
since 1990. The extracted values of $G_{M}^{n}$ at Q$^2$ of 0.1 and 
0.2 (GeV/c)$^2$ are in good agreement with previous measurements of 
Anklin {\it et al.}~\cite{Ankl94,Ankl98}. 
The Jefferson Lab data \cite{xu} provide the first precision 
information on $G^n_M$ using
a fundamentally different experimental approach than 
previous deuterium experiments.
The extracted $G^n_M$ values~\cite{e95001_prc} from the same experiment 
based on the PWIA calculation~\cite{salme} are also shown.  
While limitations exist in such an approach due to theoretical
uncertainties, the results
are in very good agreement with the recent deuterium ratio 
measurements from Mainz~\cite{Ankl98,kub01}, and in disagreement 
with results by Bruins {\it et al.}~\cite{Brui95}. 
 
Enormous progress has been made in extracting $G^n_M$ from inclusive 
quasielastic transverse asymmetry $A_{T'}$ measurement in the low 
$Q^2$ region due to the new precision data and recent
advances in three-body calculations.  
On the other hand, $A_{TL'}$ from
quasielastic $^{3}\vec{\rm He}(\vec{e},e')$ process at low $Q^2$ 
($Q^{2} \le 0.3 (GeV/c)^2$) is dominated by the proton contribution 
largely because of the smallness of $G^{n}_{E}$
and the small non-$S$ wave
part of the $^{3}$He ground state wave function. Thus, it is unreliable to
extract information on $G^{n}_{E}$ at low $Q^{2}$ from the inclusive  
$^{3}\vec{He}(\vec{e},e')$~\cite{ole} process. 
It is possible to go to higher $Q^{2} \ (Q^{2} \ge 0.3 (GeV/c)^2)$ to extract 
$G^{n}_{E}$ with respectable accuracy from quasielastic 
$^{3}\vec{\rm He}(\vec{e},e')$ measurement where the proton contribution to 
$A_{TL'}$ is under better control. It is more advantageous
to determine the neutron electric 
form factor from the $^{3}\vec{\rm He}(\vec{e},e'n)$ reaction 
rather than the $^{3}\vec{He}(\vec{e},e')$ reaction 
because the proton contribution to the asymmetry is minimized.

The experimental asymmetry for the 
coincident reaction $^{3}\vec{He}(\vec{e},e'n)$ at the 
quasielastic kinematics can be expressed as follows in PWIA:

\begin{eqnarray}
 A & = & - P_e~P_n~D  {{2 \sqrt{\tau(\tau+1)} \tan(\theta/2) G_E^n 
G_M^n \sin(\theta^\ast) \cos(\phi^\ast)}\over {{G_E^n}^2 + {G_M^n}^2 (\tau
+ 2 \tau (1+\tau) \tan^2(\theta/2))}}~~-  \nonumber \\
   &   & P_e~P_n~D   {{2 \tau\sqrt{1+\tau+(1+\tau)^2 
\tan^2(\theta/2)} \tan(\theta/2) {G_M^n}^2 
\cos(\theta^\ast)}\over {{G_E^n}^2 + {G_M^n}^2 (\tau
+ 2 \tau (1+\tau) \tan^2(\theta/2))}}   
\label{eqn:gen}
\end{eqnarray}
Here $P_e$ is the electron polarization, $P_n$ is the neutron 
polarization, $D$ is an overall dilution factor which contains
dilution from (possible) unpolarized neutrons in the target and
dilution from background neutrons generated in $(p,n)$ reactions, e.g. in
shielding walls. Eqn.~\ref{eqn:gen}
shows the obvious sensitivity to G$_E^n$ in the longitudinal-transverse
interference term. Therefore, by aligning the target spin perpendicular
to ${\vec {q}}\  (A=A_{perp})$, i.e. choosing $\theta^\ast$ = 90$^\circ$,
and $\phi^{\ast}$ = 0$^\circ$ the above equation can be rewritten
in the following form (${G^{n}_{E}}^2 \approx 0$):
\begin{equation}
G_E^n = - {A_{perp}\over{P_e P_n D}}\cdot {{G_M^n (\tau + 2\tau
(1+\tau) \tan^2(\theta/2))}\over {2 \sqrt{\tau (1+\tau)}
\tan(\theta/2)}}
\end{equation}

Aligning the target spin parallel to $\vec{q}\  (A=A_{long})$ 
reduces Eqn.~\ref{eqn:gen} to 
(${G^{n}_{E}}^2 \approx 0$):

\begin{equation}
A_{long} = -P_{e}P_{n}D{{2\sqrt{1+\tau+(1+\tau)^{2}\tan^{2}(\theta/2)}
\tan(\theta/2)} \over {1+2(1+\tau)\tan^{2}(\theta/2)}}. 
\end{equation}

This equation is completely independent of the neutron form factors
and serves as an excellent calibration reaction. Thus, one can combine
the above two equations and obtain

\begin{equation}
G^n_{E} =  \sqrt{\tau+\tau(1+\tau)\tan^{2}(\theta/2)} 
{{A_{perp}} \over {A_{long}}} G^{n}_{M}.
\end{equation}

\begin{figure}[htbp]
\centerline{\epsfig{file=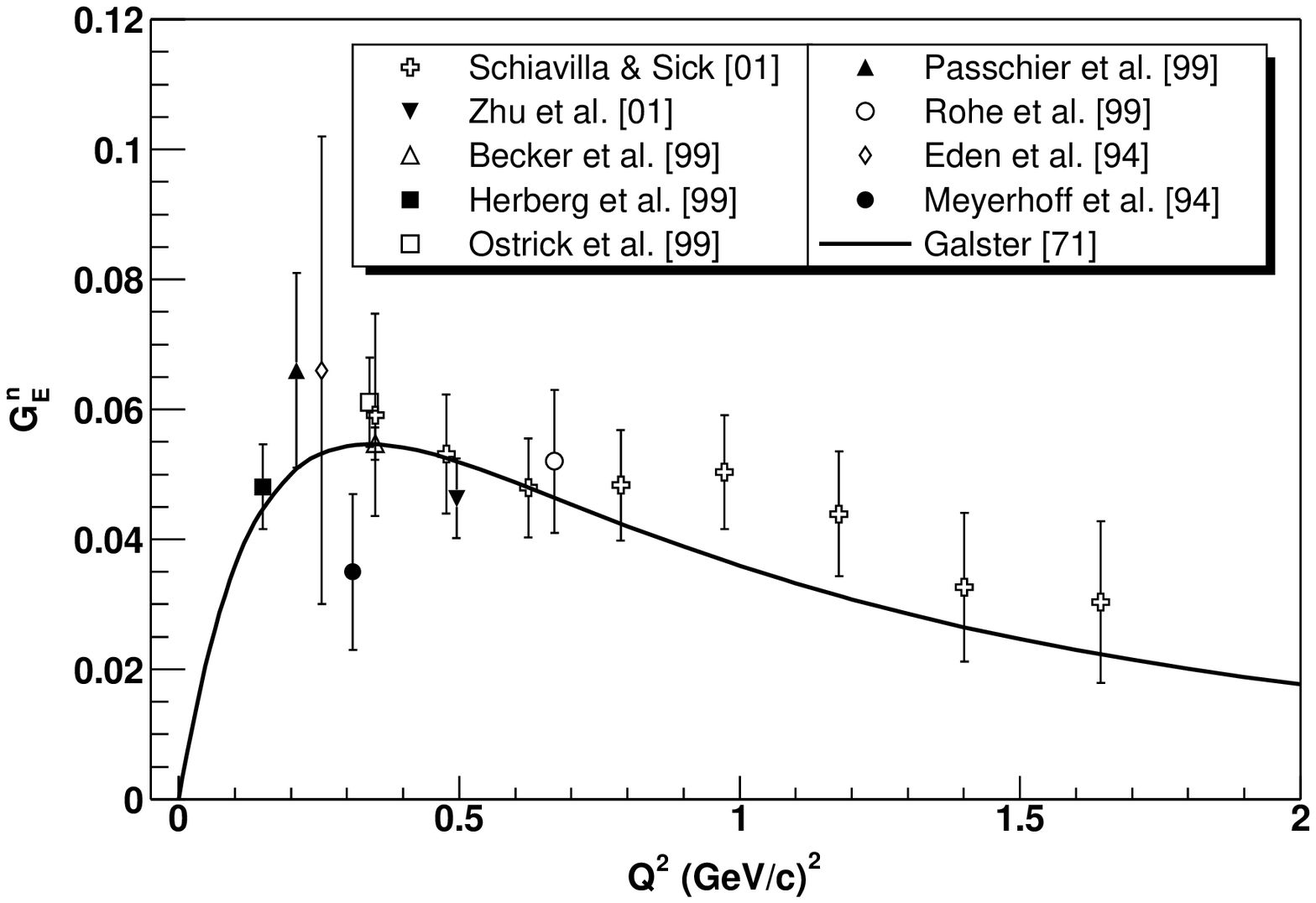,height=6.0cm,width=12.0cm}}
\fcaption{{\footnotesize\it Recent data on $G^n_E$ from polarization 
experiments. Also shown are the extracted $G^n_E$ values 
from the deuteron quadrupole form factor
data by Schivilla and Sick~\protect\cite{rocco}. The Galster 
parameterization~\cite{galster} is also shown.}}
\label{fig:genpol}
\end{figure}

Precise information on $G^n_M$ is a priori requirement before one can 
extract $G^n_E$ from such double polarization measurements. The 
PWIA picture described above is obviously over-simplified. 
To extract $G^n_E$ reliably,
corrections need to be applied, in particular the correction for the 
FSI effect, which has been carried out recently using the Faddeev 
approach~\cite{golak4,golak5}.
There is also technical difficulty involved with detecting knocked out
neutrons. Such experiments have been carried out at Mainz and NIKHEF in recent
years. The first measurement on $G^{n}_{E}$ from 
$^{3}\vec{\rm He}(\vec{e},e'n)$ was reported by Meyerhoff {\it et al.} 
\cite{Meyerhoff} in
which a high pressure polarized $^{3}$He target achieved by
the metastability-exchange optical pumping technique and the compression
method was employed. More recent Mainz measurement by 
Becker {\it et al.}~\cite{becker} and Rohe {\it et al.}~\cite{rohe}
using the same technique show much improved statistical accuracy.
The FSI correction has been applied~\cite{golak4} 
in the extracted $G^n_E$ value from the measurement by 
Becker~{\it et al.}~\cite{becker}, which is shown in Fig.~\ref{fig:genpol}. 
While the $G^n_E$ shown in Fig.~\ref{fig:genpol} 
by Meyerhoff {\it et al.}~\cite{Meyerhoff} and 
Rohe {\it et al.}~\cite{rohe} do not include corrections 
for FSI, the FSI effect is expected to be small around a $Q^2$ value of 
0.6 (GeV/c)$^2$ and the FSI effect is expected to increase the value of
$G^n_E$ by Meyerhoff {\it et al.}~\cite{Meyerhoff} to a value of around 0.05. 
Also shown in Fig.~\ref{fig:genpol} are the recent published $G^n_E$ data from
double-polarization deuterium experiments discussed previously,  
the extracted $G^n_E$ values from the deuteron quadrupole form factor
data by Schivilla and Sick~\cite{rocco}, and the Galster 
parameterization~\cite{galster}. 

\vspace{0.7in}

Enormous progress has been made in the study of the neutron 
electromagnetic form factor using polarization degrees of freedom, 
and in particular with  
spin-polarized $^3$He nuclear targets. However, it is still a very challenging 
task for both experimenters and theorists.
The non-relativistic Faddeev approach has been well tested in the low 
momentum transfer and low energy transfer regime and reliable information
on the neutron electric and magnetic form factor can be extracted in the low
$Q^2$ region.
The development of fully relativistic three-body 
calculations is required to extract the neutron electromagnetic 
form factors reliably in the relatively higher $Q^2$ region, which
is an extremely difficult task.
Recently, Kamada and collaborators~\cite{kamada} 
solved the relativistic Faddeev 
equations for the three-nucleon bound state with a Lorentz boosted 
two-nucleon potential which is introduced in the context of equal time 
relativistic quantum mechanics. Such a calculation opens the door to 
considering the relativistic Faddeev equations for three-nucleon scattering.

\section{Theoretical Progress}

While enormous experimental progress has been made 
on the subject of nucleon electromagnetic form factors, 
significant theoretical progress has also been made in recent years  
in understanding the nucleon electromagnetic structure
from the underlying theory of QCD. The newly discovered
Generalized Parton Distributions (GPDs)~\cite{Ji97,Ra96}, which can be 
accessed through deeply virtual Compton scattering and deeply virtual meson 
production, connect the nucleon form factors and the
nucleon structure functions probed in the deep-in-elastic 
scattering experiments. 
The GPDs provide new insights into the structure of the nucleon, 
and provide possibly a complete map of the nucleon wave-function.

QCD is the theory of strong interaction and has been extremely well tested 
in the high-energy region, i.e., in the perturbative QCD regime. 
Ideally, one should 
calculate the nucleon electromagnetic form factors directly from QCD in 
the non-perturbative region to confront the data. 
Unfortunately, nobody knows how to solve QCD analytically in the 
non-perturbative regime.
Lattice QCD calculations based on first principles of QCD, on the other hand, 
have shown much promise; and this field is evolving rapidly due  
both to improvements in computer architecture and to new algorithms. While 
pQCD gives prediction for the nucleon form factors in the
perturbative region, 
QCD effective theories such as the chiral perturbation theory 
can in principle provide reliable prediction in the very low energy region. 
In between the low energy region and the pQCD regime, various QCD-inspired 
models and other phenomenology models exist. 
Thus, precision data in all experimentally accessible 
region is crucial in testing these predictions. The rest of the 
section is devoted to brief 
discussions of various theoretical approaches used to calculate 
the nucleon electromagnetic form factors.

\subsection{Scaling and Perturbative QCD}

QCD is an asymptotically free theory, i.e., the strong coupling constant 
$\alpha_s \rightarrow 0$ as the inter-quark distance $\rightarrow$ 0. Thus, 
one can solve QCD using the perturbation method in the limit of 
$Q^2 \rightarrow \infty$. Such an 
approach is the so-called perturbative QCD (pQCD), 
and specific $Q^2$ dependence 
of the nucleon electromagnetic form factors can be obtained from 
the pQCD analysis.

Brodsky and Farrar~\cite{farrar} proposed the following scaling
law for the proton Dirac ($F_1$) and Pauli form factor ($F_2$) at 
large momentum transfers
based on dimensional analysis:

$$F_1 \propto ({Q^2})^{-2}, \ \ \ F_2 \sim {\frac{F_1}{Q^2}}.$$

The helicity-flipping form factor, $F_2$, is suppressed 
by ${\frac{1}{Q^2}}$ compared to the 
helicity-conserving form factor $F_1$. 
Such a prediction is a natural consequence of 
hadron helicity conservation. The hadron helicity conservation arises from
the vector coupling nature of the quark-gluon
interaction, the quark helicity conservation at high energies, 
and the neglect of the non-zero quark orbital 
angular momentum state in the nucleon.
In terms of Sach's form factors $G^p_E$ and $G^p_M$, the scaling result predicts:
${\frac{G^p_E}{G^p_M}} \rightarrow$ constant at large $Q^2$. 
Such scaling results were confirmed in a short-distance pQCD 
analysis carried out
by Brodsky and Lepage~\cite{lepage}.
Considering the proton magnetic form factor at large $Q^2$ 
in the Breit frame, the initial 
proton is moving in the $z$ direction and is struck by a highly virtual photon
carrying a large transverse momentum, ${q_{\perp}}^2 = Q^2$. The form factor
corresponds to the amplitude that the composite proton absorbs the virtual photon and stays intact.
Thus, the form factor
becomes the product of the following three probability amplitudes: 
(i) the amplitude for finding
the valence $|qqq>$ state in the incoming proton; 
(ii) the amplitude for this quark state to scatter
from the incoming photon producing the final three-quark state 
with collinear momenta and 
(iii) the amplitude for the final three-quark state to reform a proton. 
The contribution of the proton Fock states other than the valence $|qqq>$ quark 
state to the form factor is suppressed as $Q^2 \rightarrow \infty$ because
each additional constituent contributes a factor of ${\frac{\alpha_s(Q^2)}{Q^2}}$ to 
the amplitude in (ii).
 
Based on this picture, Brodsky and Lepage obtained the following 
result within their short-distance
pQCD analysis~\cite{lepage}:

\begin{eqnarray}
G_M(Q^2) &=& {\frac{32\pi^2}{9}}{\frac{\alpha_{s}^2(Q^2)}{Q^4}}\sum_{n,m}
{b_{nm}(\ln{\frac{Q^2}{\Lambda^2}})}^{-\gamma_n-\gamma_m}\left[1+O(\alpha_s(Q^2),m^2/Q^2)\right]\nonumber \\
& \rightarrow & {\frac{32\pi^2}{9}}C^2 {\frac{\alpha_{s}^2(Q^2)}{Q^4}}
{(\ln{\frac{Q^2}{\Lambda^2}})}^{-4/3\beta}(-e_{\|}),
\end{eqnarray}

\begin{figure}[htbp]
\centerline{\epsfig{file=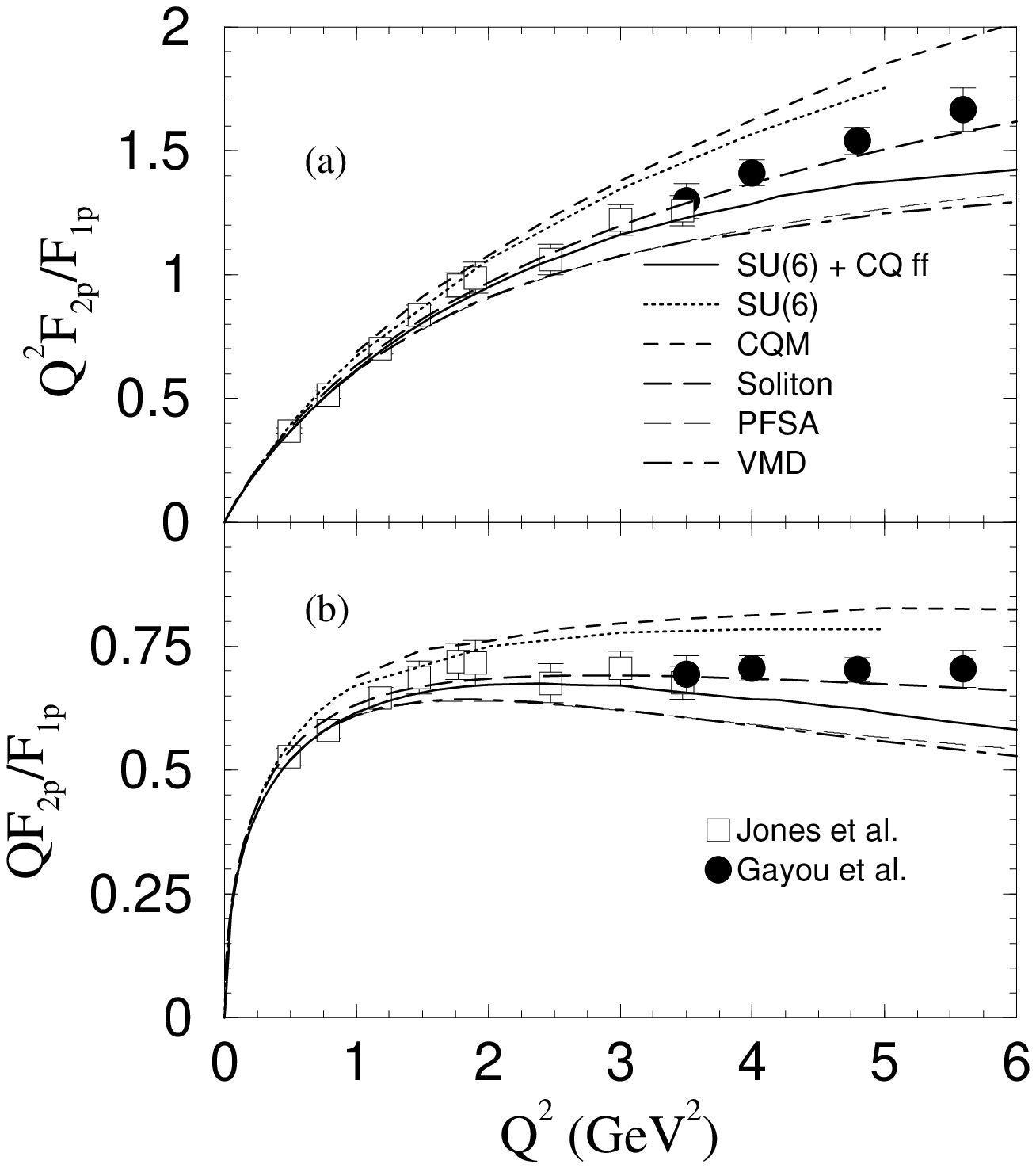,height=12.0cm,width=10.0cm}}
\fcaption{{\footnotesize\it The scaled proton Dirac and Pauli form 
factor ratio: ${\frac{Q^2F^p_2}{F^p_1}}$ (upper panel) 
and ${\frac{QF^p_2}{F^p_1}}$ (lower panel) 
as a function of $Q^2$ in (GeV/c)$^2$. The data are from Jones 
{\it et al.}~\cite{mjones} and Gayou {\it et al.}~\cite{gayou} shown 
with statistical uncertainties only.}}
\label{fig:f1f2}
\end{figure}

where $\alpha_{s}(Q^2)$ and $\Lambda$ are the QCD strong coupling constant 
and scale parameter, $b_{nm}$ and $\gamma_{m,n}$ are QCD anomalous 
dimensions and constants, and $e_{\|}$ (-$e_{\|}$) is the mean total 
charge of quarks 
with helicity parallel (anti-parallel)
to the nucleon's helicity. For protons and neutrons

$$e^p_{\|} =1, -e^p_{\|}=0, e^n_{\|}=-e^n_{\|} = -{\frac{1}{3}},$$

based on the fully symmetric flavor-helicity wave function.
While the constants $b_{nm}$ and $C$ are generally unknown for baryons, they are equal for protons
and neutrons using isospin symmetry. 
For the proton electric form factor, one obtains similar result for
the $Q^2$ dependence in the $Q^2 \rightarrow \infty$ limit and as such the short-distance 
pQCD analysis predicts the same scaling law as the dimensional analysis for the proton 
form factors:
${\frac{G^p_E}{G^p_M}} \rightarrow$ constant and 
${\frac{Q^2F_2}{F_1}} \rightarrow $ constant.
Recently, Belitsky, Ji and Yuan~\cite{belitsky}
performed a perturbative QCD analysis of the nucleon's Pauli 
form factor $F_2(Q^2)$ in the asymptotically large $Q^2$ limit. They find that
the leading contribution to $F_2(Q^2)$ goes like $1/Q^6$, which is 
consistent with the scaling result obtained by Brodsky and 
Farrar~\cite{farrar}.
The recent Jefferson Lab data~\cite{mjones,gayou} 
on the proton electric and magnetic 
form factor ratio, 
${\frac{G^p_E}{G^p_M}}$ up to a $Q^2$ value of 5.5 (GeV/c)$^2$ show a strong $Q^2$ dependence in
the ratio: the electric form factor falls off much faster than the magnetic form factor. 
In the naive picture, such $Q^2$ dependence may suggest that the proton charge 
is distributed over a larger spatial region than its counterpart, the magnetization. 
Fig.~\ref{fig:f1f2} shows data on the scaled proton Dirac and Pauli form factor ratio
${\frac{Q^2F^p_2}{F^p_1}}$ (upper panel) and ${\frac{QF^p_2}{F^p_1}}$ (lower panel) 
from Jefferson Lab experiments~\cite{mjones,gayou} 
as a function of $Q^2$ together with various predictions.
The dash-dotted curve is a new fit based on an improved vector meson 
dominance model (VMD) by Lomon~\cite{lomon1}. 
The thin long dashed curve is a point-form spectator approximation (PFSA) prediction
of the Goldstone boson exchange constituent quark model (CQM)~\cite{rfw}.
The solid and the dotted curves are the CQM calculations by 
Cardarelli and Simula~\cite{simula3} including SU(6) 
symmetry breaking with and without constituent quark form factors, 
respectively.
The long dashed curve is a relativistic chiral soliton model 
calculation~\cite{holzwarth1}. 
The dashed curve is a relativistic CQM calculation by Frank, Jennings, and Miller~\cite{frank}.

While the short-distance pQCD analysis~\cite{lepage}
predicts a constant behavior for the ${\frac{Q^2F^p_2}{F^p_1}}$ in the 
$Q^2 \rightarrow \infty$ limit, the data are in better agreement with
the ${\frac{QF^p_2}{F^p_1}}$ scaling behavior.
The data could imply that the asymptotic pQCD scaling region has not 
been reached or that hadron helicity is not conserved in the experimentally tested 
regime so far. Such an experimental observation is consistent with studies 
from polarized deuteron photodisintegration~\cite{krishni},  
polarized neutral pion production 
$^1 H(\vec{\gamma},\vec{p})\pi^0$~\cite{krishni2}, 
and deuteron tensor polarization measurements $e d \rightarrow e\vec{d}$~\cite{abbott}. 
However, Brodsky, Hwang and Hill~\cite{hwang} were able to fit 
the Jefferson Lab data using a form consistent with pQCD analysis and 
hadron helicity conservation by taking into account higher twist contributions.
Belitsky, Ji and Yuan~\cite{belitsky} also suggest that 
one should take into account
the leading non-vanishing contribution in looking at the pQCD 
scaling behavior between $F^p_1$ and $F^p_2$ form factors.  
On the other hand, Ralston and Jain~\cite{ralston} 
argue that the ${\frac{QF^p_2}{F^p_1}}$ scaling behavior is expected from pQCD when 
one takes into account contributions to the proton quark wave function from 
states with non-zero orbital angular momentum.

Following an earlier work by Frank, Jennings and Miller~\cite{frank}, 
Miller~\cite{miller2002} recently used light front dynamics 
in modeling the nucleon as a
relativistic system of three bound constituent quarks surrounded 
by a cloud of pions. While the pion 
cloud is important for understanding the nucleon structure 
at low momentum transfers, 
particularly in understanding the neutron electric form factor, 
quark effects dominate 
at large momentum transfers. Such a model is able to 
reproduce the observed 
constant behavior of $\frac{QF_2}{F_1}$ as a function 
of $Q^2$ and the $\frac{QF_2}{F_1}$ 
ratio is predicted to be a constant up to a $Q^2$ value of 20 (GeV/c)$^2$. 

\subsection{Lattice QCD Calculations}

\begin{figure}[htbp]
\centerline{\epsfig{file=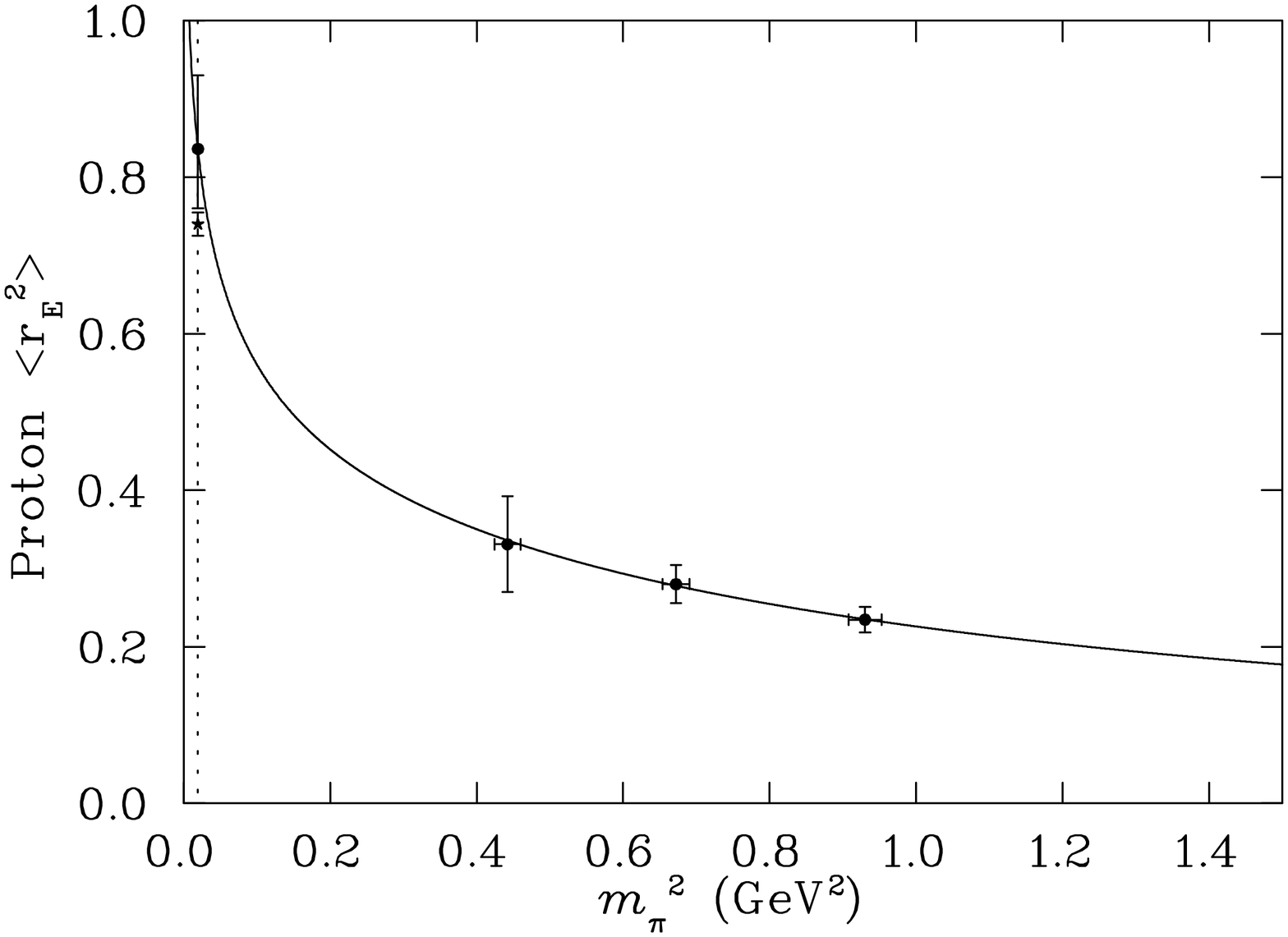,angle=90,
height=6.0cm,width=10.0cm}}
\fcaption{{\footnotesize\it Fit to the lattice QCD data for the square 
of the proton charge radius as a function of pion mass squared based 
on chiral extrapolation \protect\cite{thomas3}. The extrapolated value 
at the physical pion mass is shown by the solid dot with the large 
error bar, and the star indicates the more favorably accepted 
experimental value \cite{simon}.}}
\label{fig:lattice2}
\end{figure}

The non-perturbative nature of QCD at large distances, i.e. low momentum transfers, prevents
the analytical approach in solving QCD.
However, important conceptual and technical progress has been made over the years, especially in
the last decade or so in solving QCD on the lattice. 
While full QCD calculations have been carried out in some cases, 
most of the lattice results obtained so far were carried out
in the so-called quenched approximation in which the quark loop contributions, 
i.e. the sea quark contributions, are suppressed. Furthermore, at present and in the
foreseeable future, the lattice calculations are
only practical using rather large quark masses because of the limitations of the currently 
available computational power. As such, uncertainties in extrapolating lattice results to 
the physical quark mass are rather large, particularly with the naive linear 
extrapolation in quark mass. Thus, the challenge is to find an accurate and reliable way of 
extracting the lattice results to the physical quark mass. The extrapolation 
methods which incorporate the model independent constraints of chiral 
symmetry~\cite{thomas1}, particularly the
leading non-analytic (LNA) behavior of chiral perturbation theory~\cite{thomas2},  
and the heavy quark limit~\cite{shifman} are exciting recent developments. 
Recently, such an approach has been used~\cite{thomas} in extrapolating the lattice
result on the proton root-mean-square (rms) charge radius to the physical 
pion mass, and at present the extrapolated
value is in good agreement with the more favorably accepted 
experimental value~\cite{simon}.
Currently, lattice calculations of the nucleon form factor 
with smaller quark masses are underway~\cite{thomas3}. 
The anticipated new results together with the improved LNA 
extrapolation methods will provide more reliable lattice results 
on the nucleon form factors in the near future.

\subsection{Other Calculations of Nucleon Electromagnetic Form Factors}

\vspace{0.2in}
\begin{figure}[htbp]
\centerline{\epsfig{file=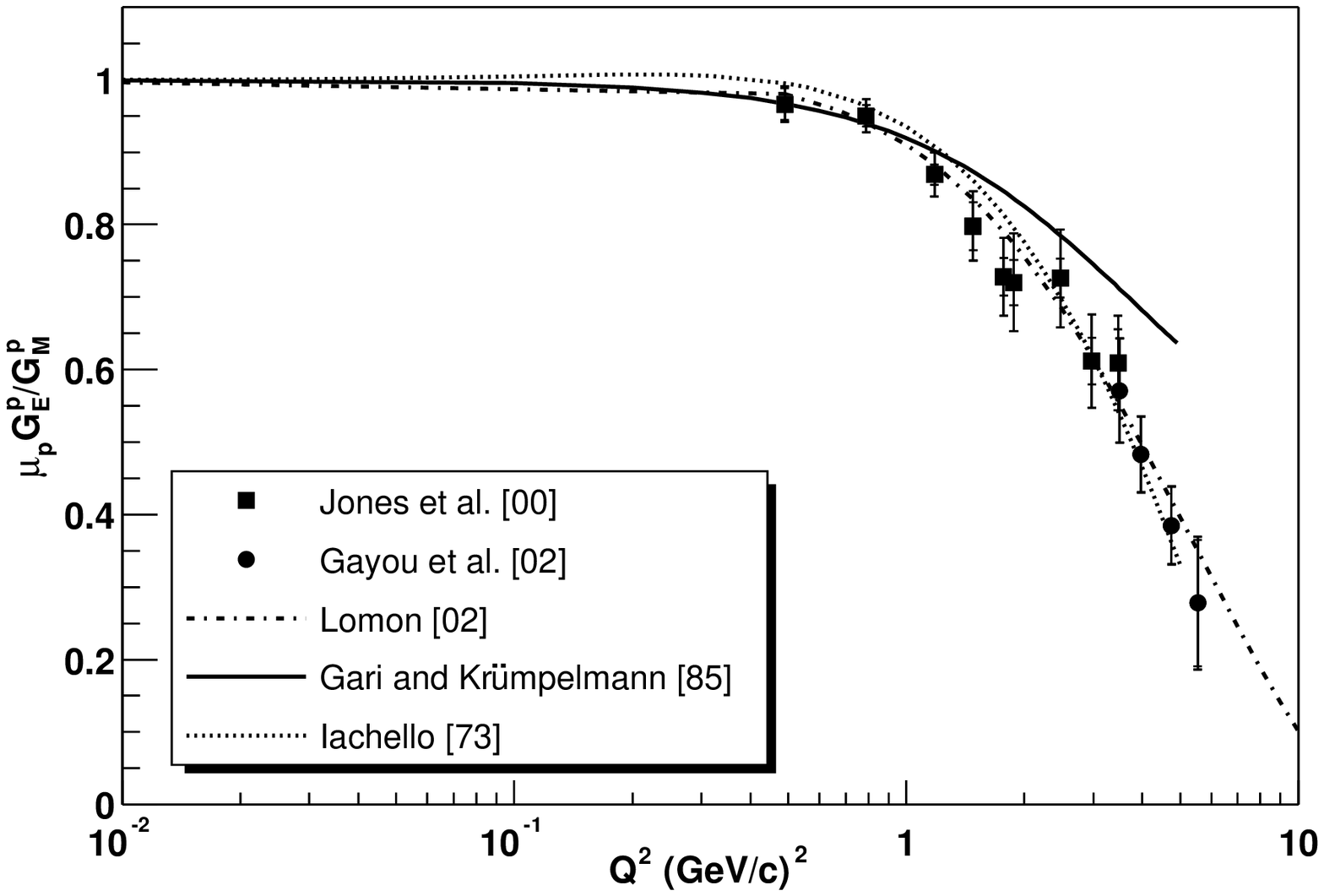,height=7.0cm,width=12.0cm}}
\fcaption{{\footnotesize\it Jefferson Lab data~\cite{mjones,gayou} 
on the proton form factor ratio, 
${\frac{\mu G^p_E}{G^p_M}}$ as a function of $Q^2$ together with 
calculations from various vector meson dominance 
models~\cite{iachello,gari1,lomon2}.}}
\label{ep_vmd}
\end{figure}

In this section, we review various models and calculations of the 
nucleon electromagnetic form factors.
While direct comparison between data on proton and neutron
electromagnetic form factors and model predictions is important,
we will restrict the discussion to those calculations which 
agree reasonably well with the trend of the recent Jefferson Lab 
data~\cite{mjones,gayou} on the proton form factor ratio and then look 
closely to see how well they describe individual form factors. 
The reason is that the Jefferson Lab data can already eliminate
some calculations based on the high $Q^2$ behavior, thus limiting
the number of candidate calculations. In this spirit, we display the
calculations and the Jefferson Lab data versus $\ln(Q^2)$ which, while
showing the Jefferson Lab data and the agreement (or lack thereof)
of the various models, accentuates the low $Q^2$ region where
lattice QCD and chiral perturbation theory are promising in making 
reliable predictions.

\smallskip

One of the earlier attempts to describe the proton form factors was
a semi-phenomenological fit by Iachello and co-workers~\cite{iachello} 
based on a model in which the scattering
amplitude is written as an intrinsic form factor of a bare nucleon
multiplied by an amplitude derived from the interaction with the 
virtual photon via vector meson dominance (VMD). Various forms of the
intrinsic bare nucleon form factor were used: dipole, monopole, eikonal.
However, since this function is multiplicative, it cancels out in
the ratio $G^{p}_E/G^{p}_M$. The VMD amplitude was written in terms of 
parameters fit to data. 
Gari and Kr\" umpelmann~\cite{gari1} 
extended the VMD model to include 
quark dynamics at large $Q^2$ via perturbative QCD. 
Because of the freedom in fitting
the parameters, this model was later~\cite{gari2} able to describe a 
subsequent set of data from SLAC that reported the ratio
$\mu_{p} G^{p}_E/G^{p}_M$ growing significantly larger than 1 
at large values of $Q^2$~\cite{rock}. 
Lomon~\cite{lomon1} extended the Gari and 
Kr\" umpelmann model~\cite{gari1} by including the width of the $\rho$ meson 
and the addition of higher mass vector meson exchanges. Such a model has been
extended further~\cite{lomon2} to include the $\omega$'(1419) isoscalar 
vector meson pole in order to describe the new Jefferson Lab 
proton form factor ratio data~\cite{mjones,gayou}. 
Fig.~\ref{ep_vmd} shows the new proton form factor ratio data 
as a function of $Q^2$ together with predictions from various VMD models 
discussed above. While these models have little absolute predictive power,
once the high $Q^2$ data have fixed the parameters, the approach
to low $Q^2$ is highly constrained. One can clearly observe that there
is significant difference between these calculations in the low $Q^2$ domain.

\smallskip

\begin{figure}[htbp]
\centerline{\epsfig{file=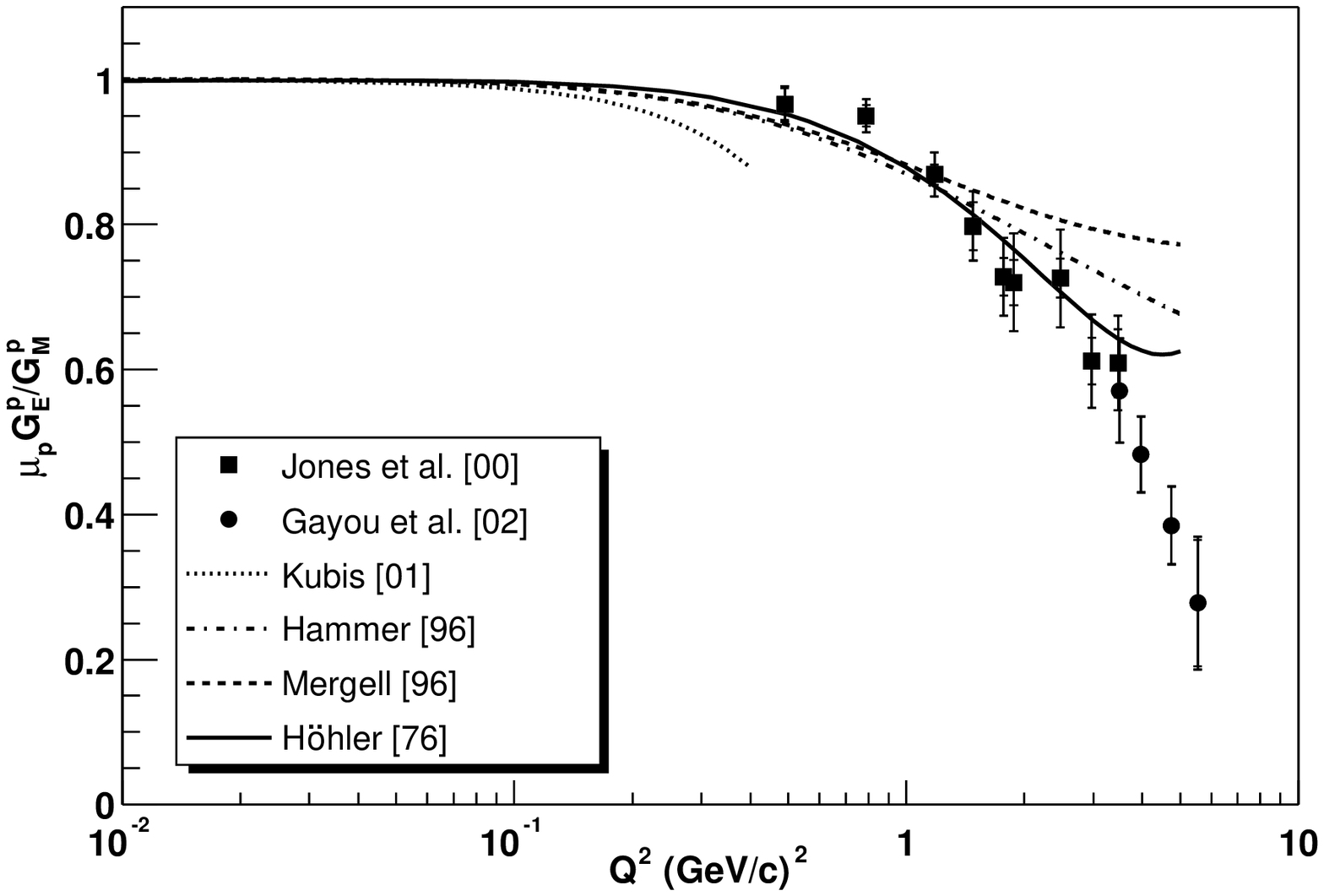,height=7.0cm,width=12.0cm}}
\fcaption{{\footnotesize\it Jefferson Lab data~\cite{mjones,gayou}
on the proton form factor ratio, ${\frac{\mu G^p_E}{G^p_M}}$ 
as a function of $Q^2$ together with 
calculations from dispersion theory fits 
\cite{hoehler,mergell,hammer,kubis}.}}
\label{ep_theory_disp}
\end{figure}

H\" ohler {\it et al.}~\cite{hoehler} fit a dispersion ansatz to 
$e-N$ scattering data. VMD contributions from $\rho$, $\omega$, $\phi$, 
$\rho'$, and $\omega'$ were included and parametrized. From these fits 
to data available at the time, a ratio $\mu_{p} G^{p}_E/G^{p}_M$ is 
obtained which is in good agreement with the present Jefferson Lab data up to 
$Q^2\approx 3$ GeV$^2$, as shown in Fig.~\ref{ep_theory_disp}, 
at which point the fitted curve begins to rise 
in contradiction to the trend of the most recent data from Jefferson Lab. 
In addition, the best fit results in a value for the rms proton radius of 0.84 fm, about 
4\% lower than the currently accepted value, indicating that the low 
$Q^2$ behavior is not quite correct. In recent years, 
these VMD/dispersion
relation approaches have been extended to include chiral perturbation 
theory~\cite{mergell,hammer,meissner1,meissner2,kubis}. 
 Mergell {\it et al.}~\cite{mergell} obtained a best fit that gave an rms 
proton radius near 0.85 fm, closer to the accepted value of 0.86 fm, 
but could not do better when simultaneously fitting the neutron data. 
Hammer {\it et al.}~\cite{hammer} included the available data on the 
form factors in the time-like region in the fit to determine the model 
parameters. These fits, also shown in Fig.~\ref{ep_theory_disp}, tend 
to under-predict the Jefferson Lab results at
low $Q^2$ and over-predict at higher $Q^2$, not being able to account
for the slope. The latest work by Kubis~\cite{kubis} which was 
restricted to the low $Q^2$ domain of 0 -- 0.4 (GeV/c)$^2$,
used the accepted proton rms radius of 0.86 fm as a constraint. The 
available data between $Q^2$ of 0.05 (GeV/c)$^2$ and 0.4 (GeV/c)$^2$ do not 
provide a severe test of the model, since they are limited by the uncertainties in the
measurements of $G^{p}_M$ in this region. However, Kubis' results show
a marked departure from the trend of the Jefferson lab data as $Q^2$
increases, decreasing far too rapidly as shown in Fig.~\ref{ep_theory_disp}.
This is not unexpected since their region of validity was claimed to
be for $Q^2 \le 0.4$ (GeV/c)$^2$.

\vspace{0.3in}
 
Several authors have calculated the proton electric and magnetic form
factors using various versions of the constituent quark model (CQM).
Chung and Coester~\cite{chung}, Aznauryan \cite{aznauryan}, and 
Schlumpf~\cite{schlumpf} all used
a relativistic CQM (RCQM) to calculate nucleon form factors in the 
$Q^2$ range of 0 - 6 (GeV/c)$^2$. Chung and Coester~\cite{chung} and 
Aznauryan~\cite{aznauryan} were able to reproduce the then available 
data on $F^{p}_1(Q^2)$ 
and $F^{p}_2(Q^2)$ in the $Q^2$ range up to about 2 - 4 GeV$^2$, above 
which the model deviated from the data particularly for Aznauryan. 
Schlumpf~\cite{schlumpf} produced good agreement with later SLAC 
data showing a rise
in the ratio of $\mu_{p} G^{p}_E/G^{p}_M$ which has now been 
shown to be contradictory with
the recent Jefferson Lab results. More recent calculations have been
made using the CQM in light front dynamics (LFCQM) by 
Cardarelli, Simula, Pace, and Salm\` e~\cite{simula1,simula2,simula3}. 
This approach
uses a one-body current operator with phenomenological form factors for
the CQs and light-front wave functions which are eigenvectors of a mass
operator which reproduces a large part of the hadron spectrum. 
The $SU(6)$ symmetry breaking effects with and without the 
constituent quark form factor are also included.
Their calculations reproduce the trend of the Jefferson lab 
results~\cite{mjones,gayou} reasonably well.

Frank, Jennings and Miller~\cite{frank} have calculated 
the proton form factors in the CQM (in light front dynamics using
the model of Schlumpf~\cite{schlumpf}) with the
primary focus of investigating the medium modifications in real nuclei.
Their results for the free proton are in reasonable
agreement with the data to the highest $Q^2$ studied so far at 
Jefferson Lab and predict a change in sign of $G^{p}_E$ at slightly
higher $Q^2$. Recently, this calculation was extended by 
Miller~\cite{miller2002}. 

  
Li\cite{li} used a relativistic quark model (RQM) in which  
symmetry is required in the center-of-mass frame. This has the effect of adding
additional terms to the baryon wave function. The original wave function
obtained before the $SU(6)$ symmetry requirement is constructed from
spinors of quarks with zero momentum. The additional terms generated
by the $SU(6)$ symmetry requirement are constructed from spinors of
anti-quarks with zero momentum. Taken together with the original terms,
these represent the inclusion of the sea quarks. The results of this
calculation originally preceded the publication of the Jefferson Lab
results~\cite{mjones}, and the model gives good agreement with the data.

\begin{figure}[htbp]
\centerline{\epsfig{file=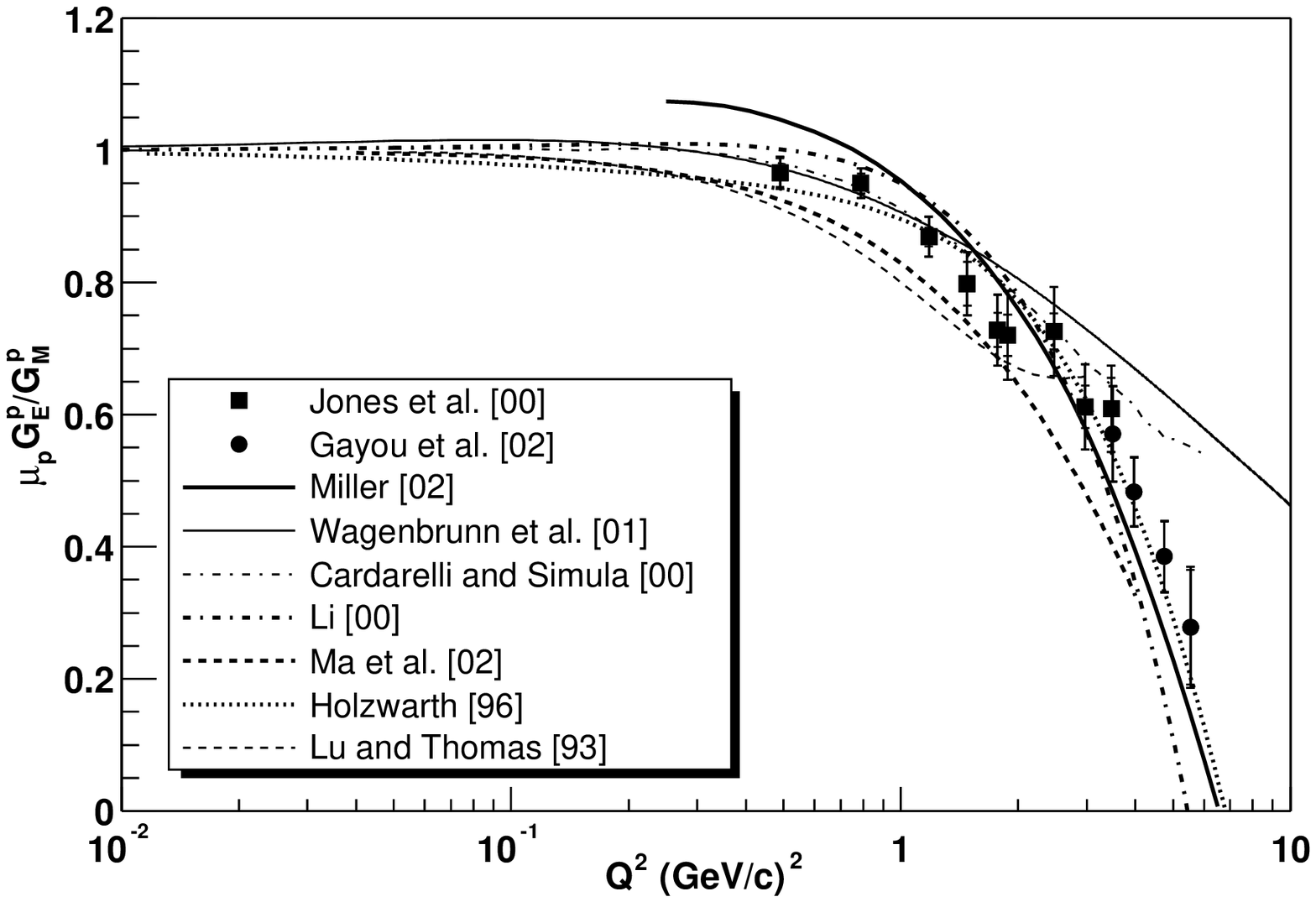,height=7.0cm,width=12.0cm}}
\fcaption{{\footnotesize\it Jefferson Lab data~\cite{mjones,gayou} 
on the proton form 
factor ratio, ${\frac{\mu G^p_E}{G^p_M}}$ as a function of $Q^2$ together with 
calculations from constituent quark 
models \cite{simula3,miller2002,holzwarth1,li,lu3,rfw,ma}.}}
\label{ep_theory_cqm2}
\end{figure}

Ma, Qing and Schmidt~\cite{ma} calculated the nucleon electromagnetic form
factors within a simple relativistic quark spectator-diquark model using the
light-cone formalism. Melosh rotations are applied to both quark and 
vector diquark. They results describe the available experimental data 
well as shown Fig.~\ref{ep_theory_cqm2}.
Recently, Wagenbrunn, Boffi, Klink, Plessas and Radici~\cite{rfw} calculated
the neutron and proton electromagnetic form factors for the first time using 
the Goldstone-boson-exchange constituent quark model. The calculations 
are performed in a covariant framework using the point-form approach to 
relativistic quantum mechanics, and are shown in Fig.~\ref{ep_theory_cqm2}.

Another recent calculation of this type is from a perturbative 
chiral quark model 
(pChQM) by Lyubovitskij {\it et al.}~\cite{Lyubovitskij}. In this model,
the effective Lagrangian describes relativistic quarks moving in a self
consistent field, the scalar component of which provides confinement while
the time component of the vector potential is responsible for short
range fluctuations of the gluon field configurations, leading to a 
cloud of Goldstone bosons ($\pi,K,\eta$), as required by chiral symmetry.
The results of these calculations drop much more rapidly than the
experimental data. They also drop off more rapidly than either the results
from the cloudy bag model (CBM), which uses a similar model, 
or those from the RCQM of Simula and co-workers.

Holzwarth has used a Skyrme/soliton model~\cite{holzwarth1} to 
calculate the proton form actors and included loop 
corrections~\cite{holzwarth2}. His results agree very well 
with the Jefferson Lab measurements. This results from a significant
deviation of $G^{p}_E$ below the ``standard'' dipole form. In fact, it is
a general feature of these models that $G^{p}_E$ changes sign for $Q^2$
somewhere larger than 5 GeV$^2$ (although exactly where is not well
predicted), while $G^{p}_M$ either does not change sign at all or does so
at a very large value of $Q^2$.  
They also agree very well with measured $G^{p}_E$ and $G^{p}_M$ values in the
low $Q^2$ region and yield a proton rms radius of 0.869 fm, in close
agreement with the currently accepted value. 

\vspace{0.3in}

\vspace{0.3in}
\begin{figure}[htbp]
\centerline{\epsfig{file=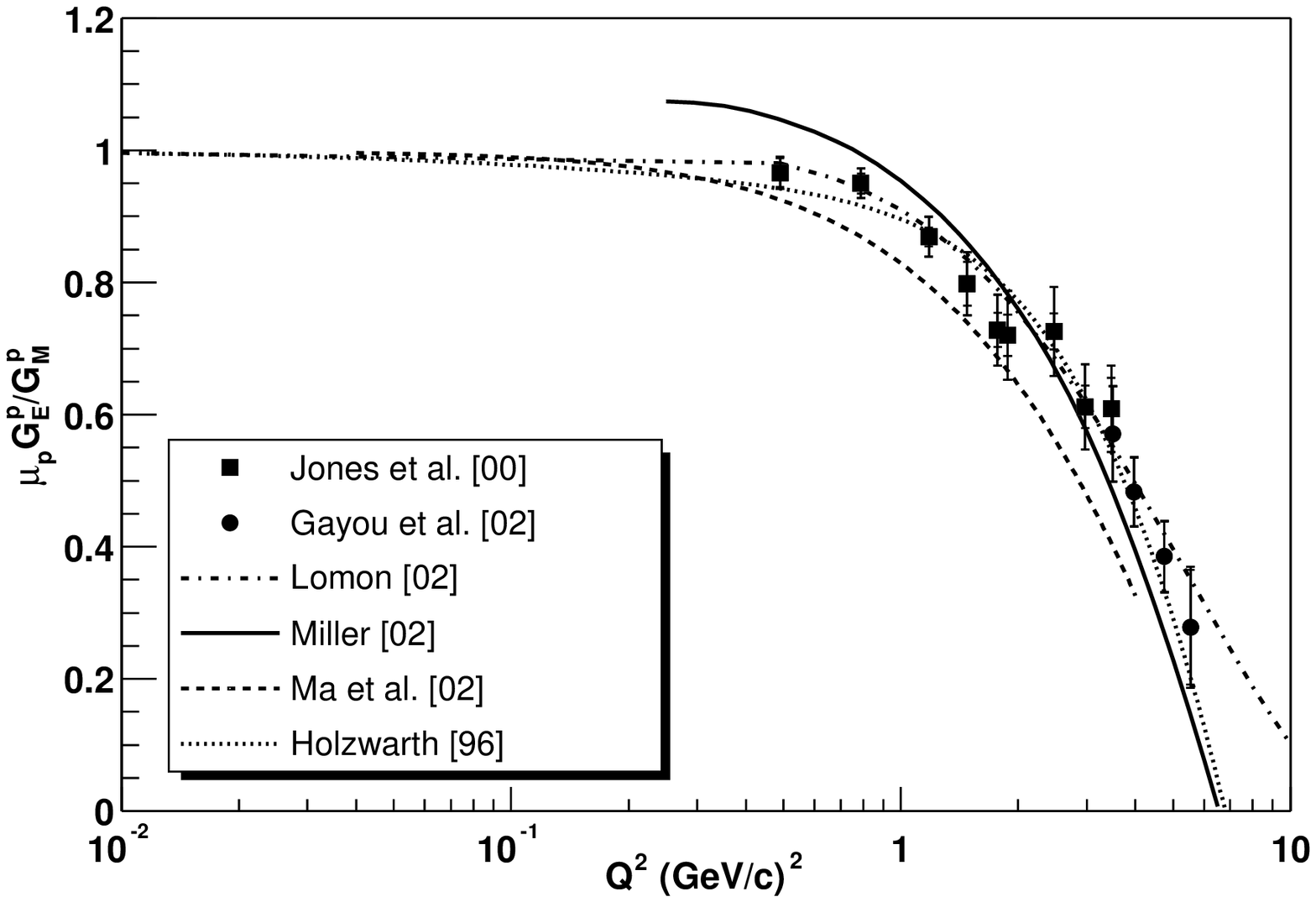,height=7.0cm,width=12.0cm}}
\fcaption{{\footnotesize\it Jefferson Lab data~\cite{mjones,gayou} 
on the proton form 
factor ratio, ${\frac{\mu G^p_E}{G^p_M}}$ as a function of $Q^2$ together with 
calculations by Lomon \cite{lomon2} (dash-dotted), Miller \cite{miller2002} 
(solid curve), Holzwarth \cite{holzwarth1} (dotted curve), 
and Ma, Qing, and Schmidt~\cite{ma} (dashed curve).}}
\label{ff_ratio_best}
\end{figure}

In the MIT Bag Model, culminating in the so-called cloudy bag model 
(CBM) which includes the bare MIT bag plus a meson cloud, quarks are
described as independent particles confined in a rigid spherical 
well. Such calculations have been performed by 
Lu and Thomas~\cite{lu1,lu2,lu3}. The introduction of 
the pion cloud improves the static properties of 
the nucleon by restoring chiral symmetry and also provides a convenient
connection to $\pi N$ and $NN$ scattering. The result of these
calculations is a ratio of $G^{p}_E/G^{p}_M$ which decreases much more 
rapidly than indicated by the Jefferson Lab data. However,
it provides a considerable improvement, as expected, over the bag model 
without the pion cloud which predicts a drop to zero at $Q^2 = 1.5$ GeV$^2$
with a subsequent change of sign.
The CBM calculation~\cite{lu3} with a bag radius of 0.7 fm is also shown in 
Fig.~\ref{ep_theory_cqm2}, and 
the abrupt change of slope between 2 and 3 GeV$^2$
is inconsistent with the newer JLab results by 
Gayou {\it et al.}~\cite{gayou}.


In summary, the Soliton 
model calculation by Holzwarth~\cite{holzwarth1}, the extended VDM model by 
Lomon \cite{lomon2}, the
relativistic CQM model calculation by Miller~\cite{miller2002}, which is 
an improvement over its earlier version~\cite{frank}, and the relativistic 
quark spectator-diquark model
calculation by Ma, Qing and Schmidt~\cite{ma},
describe the new Jefferson Lab proton form factor ratio
data very well. Fig.~16 shows the comparison between these models and the 
Jefferson Lab data~\cite{mjones,gayou}.

\begin{figure}[htbp]
\centerline{\epsfig{file=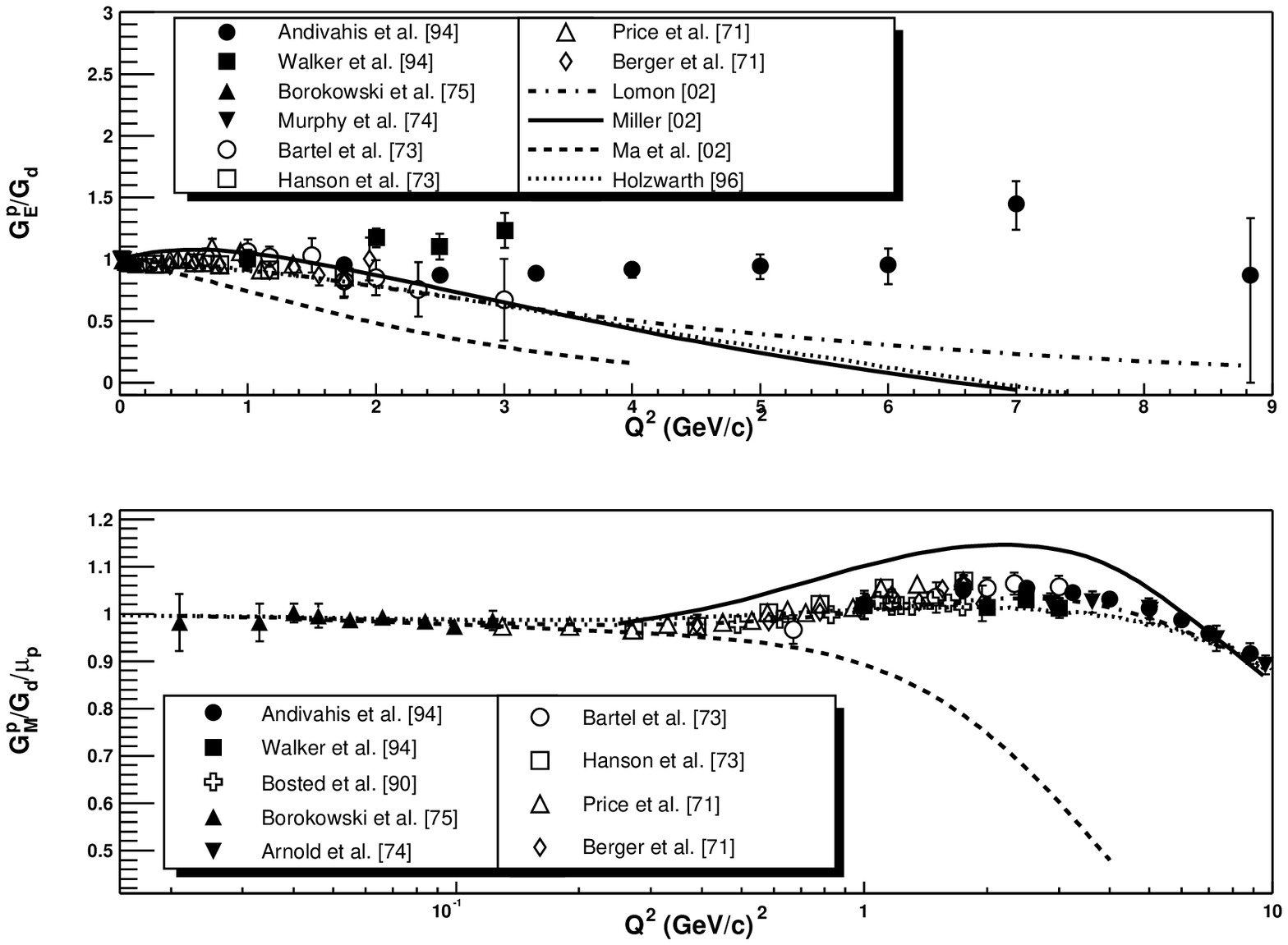,height=8.0cm,width=12.0cm}}
\fcaption{{\footnotesize\it World data on proton electromagnetic form 
factor, $G^p_E$ and ${\frac{G^p_M}{\mu_n}}$ since 1970 
in the unit of the standard dipole 
parameterization as a function of $Q^2$ together with 
calculations by Lomon \cite{lomon2} (dash-dotted), Miller \cite{miller2002} 
(solid curve), Holzwarth \cite{holzwarth1} (dotted curve), 
and Ma, Qing, and Schmidt~\cite{ma} (dashed curve).}}
\label{pff_best}
\end{figure}

While it is important for a model to explain the observed $Q^2$ 
dependence of the proton form factor ratio, more stringent tests of the model 
can be provided by comparing calculations with data on 
all four nucleon form factors 
over the entire experimental accessible momentum transfer region.
Fig.~17 shows the proton electric and magnetic form factor data 
published since 1970 obtained by the Rosenbluth method 
from unpolarized cross-section experiments, together with
predictions from these models; Fig.~18 shows the corresponding neutron data 
and the model predictions. While most of the models are in good agreement with
the proton magnetic form factor data,
the departure of the model predictions from the
$G^p_E$ data at high $Q^2$ is not surprising. These models 
describe the trend of the Jefferson Lab proton form factor ratio data well, 
which are in disagreement with the proton form factor ratio formed by 
$G^p_E$ and $G^p_M$ obtained by the Rosenbluth separation method from 
unpolarized cross-section measurements.
As shown in Fig.~18, none of these models are capable of describing 
the neutron electric and magnetic form factor data simultaneously in both
the low $Q^2$ and high $Q^2$ region. 
While enormous theoretical progress has been made in recent years, 
significantly improved theoretical understanding the nucleon electromagnetic 
form factor is urgently needed.

\smallskip

\vspace{0.3in}
\begin{figure}[htbp]
\centerline{\epsfig{file=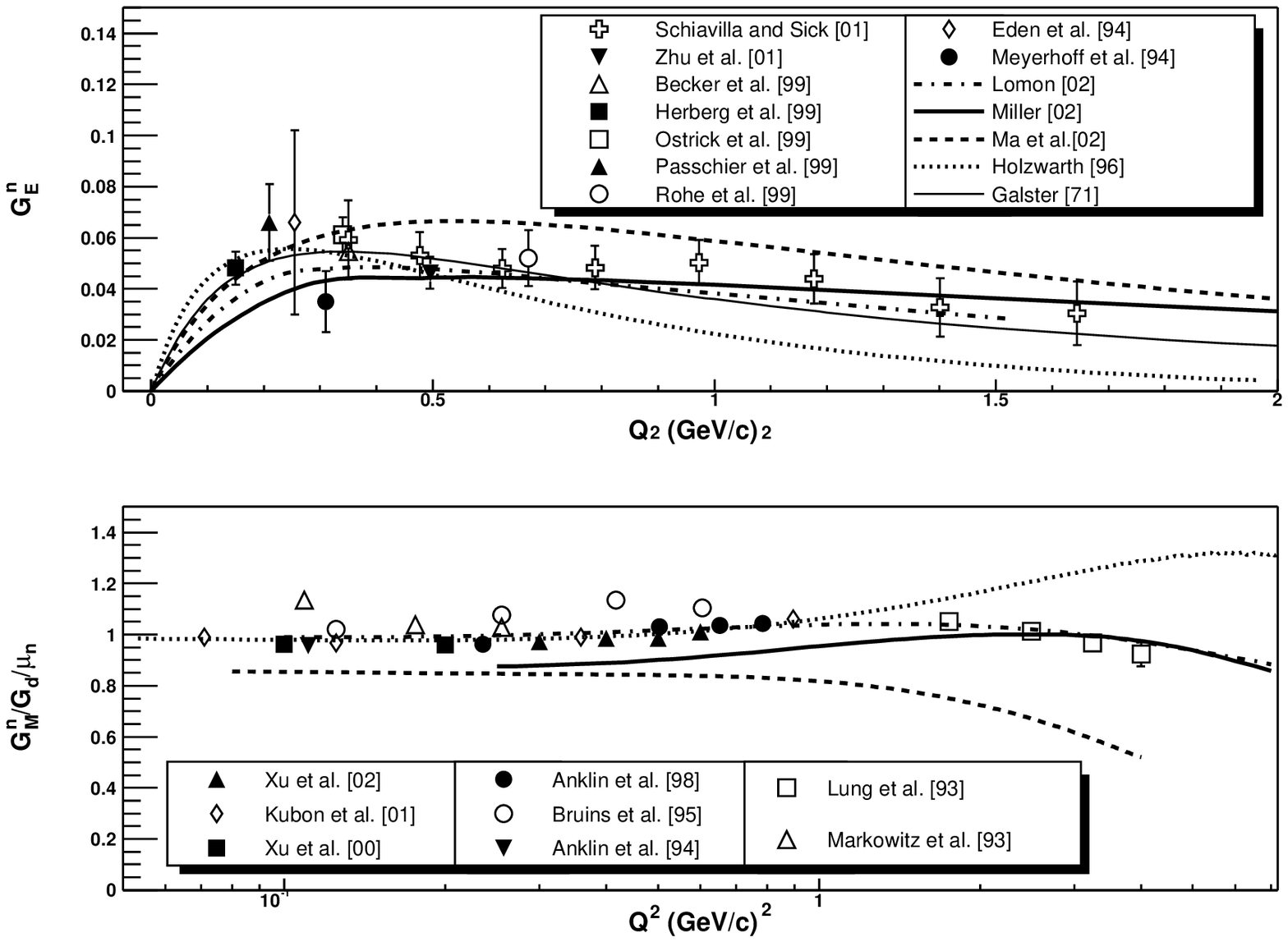,height=8.0cm,width=12.0cm}}
\fcaption{{\footnotesize\it World data on neutron electromagnetic form 
factor, $G^n_E$ and ${\frac{G^n_M}{\mu_n}}$ since 1990 
as a function of $Q^2$ together with 
calculations by Lomon \cite{lomon2} (dash-dotted), Miller \cite{miller2002} 
(solid curve), Holzwarth \cite{holzwarth1} (dotted curve), 
and Ma, Qing, and Schmidt \cite{ma} (dashed curve).}}
\label{nff_best}
\end{figure}

\section{Future Outlook}

New, high precision data have been obtained in recent years 
from different electron accelerator laboratories around the world, 
thanks to new high luminosity facilities, and novel polarization experimental 
techniques. Future measurements extending into even higher 
momentum transfer region will be carried out in the near future, 
particularly with
the planned energy upgrade of CEBAF to 12 GeV at Jefferson 
Lab~\cite{12gevwhitepaper}. With this upgrade, the proton form factor 
ratio measurement can be extended to a $Q^2$ value of 14 (GeV/c)$^2$~\cite{saha}. 
Such an extension will provide more stringent tests of 
various pQCD approaches discussed above.
Furthermore, the neutron electric form factor measurement 
can be extended to a $Q^2$ 
value of 4 (GeV/c)$^2$ using the $\vec{^{3}He}(\vec{e},e'n)$ 
reaction~\cite{bogdan}, and the neutron magnetic 
form factor measurement can be extended to 
a $Q^2$ value of about 10 (GeV/c)$^2$ 
employing the ratio technique of D(e,e'n)/D(e,e'p)~\cite{brook}. 
This technique has been employed successfully recently using the CLAS
detector at Jefferson Lab \cite{brook2}.

Precision measurements of nucleon form factor in the low momentum 
transfer region, 
especially in the limit of $Q^2 \rightarrow 0$, 
which allows the determination of the nucleon
electromagnetic radii, are equally important. This is 
a region where lattice QCD is 
anticipated to make reliable predictions in the near future 
and where effective field 
theories are expected to work. 
At MIT-Bates, 
the newly constructed BLAST detector will determine
the nucleon electromagnetic form factor with high precision 
in the $Q^2 \le 1.0 
(GeV/c)^2$ region using the novel experimental technique of scattering 
longitudinally polarized electrons
in an electron storage ring from polarized internal gas targets with a large 
acceptance detector.
The proton rms charge radius will be determined with much 
more higher precision from BLAST than from existing measurements 
from unpolarized electron-proton elastic 
scattering experiments.
Such information will be sufficiently precise
to allow high precision tests of QED from measurements of the hydrogen
Lamb shift and equally importantly, will provide reliable tests of
Lattice QCD calculations.

\section{Acknowledgements}

The author is
 grateful to many people for their help in the 
process of writing this paper. 
She would like to thank Drs. G. Holzwarth, X. Ji, B.A. Li, K.F. Liu, 
E. Lomon, B.Q. Ma, G.A. Miller, W. Plessas, I. Schmidt,
S. Simula, A.W. Thomas, and R. Wagenbrunn for 
helpful discussions and for providing their calculations. 
She would like to thank her experimental colleagues: 
Drs. J.R. Calarco, O. Gayou, 
D. Higinbotham, R. Madey, A. Semenov, and I. Sick 
for their help. She is 
particularly indebted to Jason Seely for his tireless effort in
maintaining the nucleon form factor data base, for making most of the 
figures in this article, and for careful reading of the manuscript.
Finally, she thanks Dr. E.M. Henley for  
his valuable comments and suggestions.
This work is supported   
by the U.S. Department of Energy under 
contract number DE-FC02-94ER40818 and DE-FG02-03ER41231. The author 
also acknowledges the Outstanding Junior Faculty Investigator Award 
(OJI) in Nuclear Physics from the U.S. Department of Energy.

\textheight=7.8truein
\setcounter{footnote}{0}
\renewcommand{\thefootnote}{\alph{footnote}}

\end{document}